\documentclass[aps,prb,floatfix,twocolumn,superscriptaddress,showpacs,amsmath,amssymb]{revtex4}
\usepackage{graphicx}
\usepackage{epstopdf}
\usepackage{epsfig} 
\usepackage{esint} 
\allowdisplaybreaks

\newcommand{\be}{\begin{equation}}
\newcommand{\ee}{\end{equation}}
\newcommand{\bea}{\begin{eqnarray}}
\newcommand{\eea}{\end{eqnarray}}
\newcommand{\ba}{\begin{eqnarray*}}
\newcommand{\ea}{\end{eqnarray*}}
\newcommand{\dagga}{{\phantom{\dagger}}}
\newcommand{\bR}{\mathbf{R}}
\newcommand{\bQ}{\mathbf{Q}}

\newcommand{\bq}{\mathbf{q}}
\newcommand{\bqp}{\mathbf{q'}}
\newcommand{\bk}{\mathbf{k}}

\newcommand{\bkp}{\mathbf{k'}}
\newcommand{\bp}{\mathbf{p}}

\newcommand{\br}{\mathbf{r}}
\newcommand{\bnot}{\mathbf{0}}

\newcommand{\dis}{\displaystyle}

\newcommand{\up}{\uparrow}
\newcommand{\down}{\downarrow}
\newcommand{\fract}[2]{\frac{\dis #1}{\dis #2}}
\newcommand{\Tr}{\mathrm{Tr}}
\newcommand{\eqn}[1]{(\ref{#1})}
\newcommand{\ket}[1]{\mid\! #1\rangle}
\newcommand{\bra}[1]{\langle #1\!\mid}

\newcommand{\bep}{{\overline{\epsilon}}}
\newcommand{\ep}{{\epsilon}}

\newcommand{\Sin}{\sum^{\text{IN}}}
\newcommand{\Sout}{\sum^{\text{OUT}}}

\newcommand{\bw}{\begin{widetext}}
\newcommand{\ew}{\end{widetext}}
\newtheorem{law}{Assumption}

\begin{document}
\title{Absence of thermalization in a Fermi liquid}

\author{Anna Maraga} 
\affiliation{International School for
  Advanced Studies (SISSA), Via Bonomea  265, I-34136 Trieste, Italy} 
\author{Alessandro Silva} 
\affiliation{International School for
  Advanced Studies (SISSA), Via Bonomea  265, I-34136 Trieste, Italy} 
  \affiliation{Abdus Salam ICTP, Strada Costiera 11, 34100 Trieste, Italy}
\author{Michele Fabrizio} 
\affiliation{International School for
  Advanced Studies (SISSA), Via Bonomea  265, I-34136 Trieste, Italy}

\date{\today} 

\pacs{71.10.-w, 05.30.Fk, 05.70.Ln}

\begin{abstract}
We study a weak interaction quench in a three-dimensional Fermi gas. We first show that, 
under some general assumptions on time-dependent perturbation theory, 
the perturbative expansion of the long-wavelength structure factor $S(\bq)$ is not compatible with the hypothesis that steady-state averages correspond to thermal ones. In particular, $S(\bq)$ 
does develop an analytical component 
$\sim const. + O(q^2)$ at $\bq\to\bnot$, as implied by thermalization, but, in contrast, it maintains a non-analytic part $\sim |\bq|$ characteristic of a Fermi-liquid at zero-temperature. In real space, this 
non-analyticity corresponds to persisting power-law decaying density-density correlations, whereas thermalization would predict only an exponential decay. 
We next consider the case of a dilute gas, where one can obtain non-perturbative results in the interaction strength but at lowest order in the density.  We find that in the steady-state the momentum distribution jump at the Fermi surface remains finite, though smaller than in equilibrium, up to second order in $k_F f_0$, where $f_0$ is the scattering length of two particles in the vacuum. Both results question the emergence of a finite length scale in the quench-dynamics as expected by thermalization.  
\end{abstract}

\maketitle

\section{Introduction}
\label{Introduction}

Landau's Fermi-liquid theory,\cite{Landau-1,Landau-2} one of the milestones in the quantum theory of many-body systems, explains why interacting fermions almost always display thermodynamic and transport properties similar to 
those of non-interacting particles, which is e.g. the reason of success of the Drude-Sommerfeld description of normal metals in terms of free-electrons.

In the original paper,\cite{Landau-1} Landau developed  semi-phenomenologically his Fermi-liquid theory by postulating that the low-lying eigenstates of interacting fermions are in one-to-one 
correspondence to those of free fermions, namely that, if interaction is slowly turned on, the non-interacting low-energy eigenstates {\sl adiabatically} evolve into the fully-interacting ones. Since the former are Fock states labelled by occupation numbers $n_{\bk\sigma}$ in momentum and spin space, the same labeling can be maintained also for the fully-interacting eigenstates, in which case $n_{\bk\sigma}$ refer to so-called {\sl quasi-particles}.   

Such an adiabatic assumption might at first sight seem unjustified because of the absence of an energy gap in the spectrum. However, 
even though the single-particle spectrum of a normal Fermi gas is gapless, the density of states of particle-hole pairs, which 
are the only excitations generated in the process of switching interaction on, is 
actually power-law vanishing at low energy and in more than one dimension. This sort of pseudo-gapped behavior, though it does not rigorously prove the validity of adiabatic continuation, at least makes the latter less unlikely.

The past few years have witnessed a growing interest in quantum non-equilibrium phenomena,~\cite{Polkovnikov2011} mostly motivated by cold-atom~\cite{bloch_review} and ultrafast pump-probe spectroscopy experiments.~\cite{Orenstein} Among the issues under discussion, one of the most important is whether a macroscopic but isolated quantum system can serve as its own dissipative bath. Imagine that such a system is initially supplied with an extensive excess energy and then let evolve until it relaxes to a steady state. 
Should quantum ergodicity hold,~\cite{srednicki_94,neumann_29} the steady-state values of local observables would coincide with thermal averages at an effective temperature $T_*$ such that the internal energy  
coincides with the initial one, which is conserved during the unitary evolution. The problem under discussion is when and how the above {\sl thermalization hypothesis} holds.~\cite{rigol_07,rigol_08,Polkovnikov2011}

Experiments on physical realizations of almost integrable models
~\cite{kinoshita,Gring2012} do not actually find evidences of thermalization 
{\sl within accessible time scales}. 
The common wisdom is that such non-thermal state
behaviour, sometimes refereed to as {\sl pre-thermalization}, and observed in
a wealth of different model calculations~\cite{Essler2013,Hamerla-1d,
Marcuzzi2013,Mitra2013} 
even far from integrability,~\cite{Kehrein,Werner-PRL,Hamerla-2d,Berges2004,Cazalilla-vabbe'} will eventually give in to a thermal state at long times. However, the final flow towards thermal equilibrium remains so far rather elusive.

In the specific case of a Landau-Fermi liquid, it is reasonable to expect that thermalization does take place because of the continuum of gapless excitations that can efficiently dissipate and redistribute the excess energy. It is however evident that thermalization after an interaction quench 
is not fully compatible with the aforementioned  Landau's adiabatic assumption.~\cite{Landau-1} 
In fact, if the interaction is turned on in a finite time $\tau$, however long,  the final-state energy differs from the ground state one by an extensive amount. Should thermalization indeed occur, such a finite energy density would translate into a finite temperature 
$T_*\not =0$. Since a Landau-Fermi liquid can be regarded as a quantum-critical state where all equal-time correlation functions decay as a power-law in the distance at zero-temperature and exponentially at any $T\not =0$, this would imply that, no matter how large $\tau$ is, if thermalization holds the initial non-interacting Fermi sea will \it never \rm evolve into the interacting ground state. 
       
Therefore thermalization in a Landau-Fermi liquid is not as trivial as one could have envisaged, and indeed this issue is still open and controversial. Early calculations by Moeckel and Kehrein,~\cite{Kehrein} based on the flow equation method at second order in perturbation theory, point to the existence of a quasi-steady state resembling a zero-temperature Fermi liquid.
However, the same authors argued that such a ``pre-thermal'' regime had eventually to give up to a genuine thermal state if higher order corrections were kept into account.~\cite{Kehrein} Evidences of a pre-thermal regime were also found by Eckestein, Kollar and Werner,~\cite{Werner-PRL} who numerically simulated through dynamical mean field theory a sudden interaction-quench in the infinite-dimensional half-filled Hubbard model. In particular, they found that the energy distribution jump at the Fermi energy, after a plateau compatible with pre-thermalization of Ref.~\onlinecite{Kehrein}, starts a steady decrease at longer times. Unfortunately, the affordable simulation time was too short to ascertain whether the jump does indeed relax to zero as expected by thermalization, especially at  interaction values  safely low to exclude any influence by the Mott transition. The "pre-thermal" regime has been later employed by Stark and Kollar~\cite{Kollar} as initial condition to integrate a Boltzmann kinetic equation for the energy distribution function, on the assumption that Wick's theorem holds for quasiparticles at sufficiently low interaction-strength. The solution indeed flows to a thermal distribution, which is unsurprising given those premises. Therefore these results 
do not provide a rigorous proof that thermalization does take place, although the good agreement with dynamical mean field theory results, in the temporal range in which the latter are available, is a strong evidence in favour. Hamerla and Uhrig\cite{Hamerla-1d,Hamerla-2d} have attached the same problem by truncating at some high order in perturbation theory the Heisenberg equation of motion for the momentum distribution. At weak coupling or away from half-filling, namely when the dynamical evolution is far from being affected by the Mott transition, also their calculations do not allow to conclude without doubts that the momentum distribution jump vanishes in the steady-state, hence that the latter indeed resembles a Fermi gas at finite temperature for what concerns long-range correlations.

The above list of earlier attempts is by no means exhaustive and is only meant to convey the message that the issue of thermalization in a Fermi liquid, in the specific sense of correlations that are initially power-law in real space and turn exponential in the steady-state, is not settled yet. Here we shall contribute to this puzzle 
first formulating the question in a framework that is amenable to perturbative calculations, section \ref{Formulating the question}, 
then showing in section \ref{Second order perturbation theory results} that signals of the putative exponential decay of the steady-state correlations do not appear at the expected order in perturbation theory. In section 
\ref{Bosonization} we shall instead show that steady-state correlations seem to remain power-law even when non-perturbative results in the interaction are accessible, specifically in the dilute Fermi gas at leading order in the density.  An attempt to interpret these results is discussed in the conclusive section \ref{Conclusions}.

\section{Consequences of thermalization for a weak quench}
\label{Formulating the question}

Let us start our analysis by formulating a general criterion to test the occurrence of thermalization. Hereafter 
we shall deal with the specific example of a three-dimensional Fermi-Hubbard model $\mathcal{H}(t)=
\mathcal{H}_0 + U(t)\,\mathcal{U}$, with 
\bea
\mathcal{H}_0 &=& \sum_{\bk\sigma}\!\epsilon_\bk c^\dagger_{\bk\sigma}
c^\dagga_{\bk\sigma},
\label{Ham-0}\\
\mathcal{U} &=& \fract{1}{V}\!\!\sum_{\bk,\bp,\bq\not=\bnot}\!\!\!
c^\dagger_{\bk\up}c^\dagger_{\bp+\bq\down}c^\dagga_{\bp\down}c^\dagga_{\bk+\bq\up},\;\;\;\:
\label{Ham-U}
\eea
where $V$ is the volume and we have removed from $\mathcal{U}$ the $\bq=\bnot$ Hartree-term, which refers to conserved quantities.  
We shall further assume a low electron density $n=N/V\ll 1$, where $N$ is the number of particles, so 
that it is safe to approximate the dispersion relation with a quadratic form, $\epsilon_\bk \propto |\bk|^2$, 
and we will always take the thermodynamic limit before every other limit, $V\to \infty$ at constant 
density $n$. 

The interaction is turned on as  
\be
U(t) = U\,\Big(1-\text{e}^{-\epsilon t}\Big),\label{U(t)}
\ee
with $\epsilon>0$ and an interaction strength $U\!\ll\! T_F$, where $T_F$ is the Fermi temperature. The expression 
\eqn{U(t)} interpolates smoothly between the sudden quench, $\epsilon\to\infty$, and the adiabatic switching, first send $t\to\infty$ and later $\epsilon\to 0$. 

Imagine that the system is prepared in the Fermi sea $\ket{0}$, ground state of $\mathcal{H}_0$ with energy $E_0$, and let evolve with the full Hamiltonian $\mathcal{H}(t)=
\mathcal{H}_0 + U(t)\,\mathcal{U}$, which corresponds to a time-dependent state $\ket{\Psi(t)}$ that satisfies ($\hbar=1$) 
\be
i \,\fract{\partial \ket{\Psi(t)}}{\partial t} = \mathcal{H}(t)\ket{\Psi(t)},
\ee
with $\ket{\Psi(0)}=\ket{0}$. 
The interaction switching protocol changes the total energy of the system according to 
\bea
E_* - E_0 &=& \int_0^\infty dt \,\langle \Psi(t)\mid \fract{\partial  \mathcal{H}(t)}{\partial t}
\mid \Psi(t)\rangle \label{1:Delta E} \\
&\simeq& -V\, \fract{U^2}{4}
\int \fract{d^3q}{(2\pi)^3}\,\int_0 d\omega\, d\omega'\,\, 
\rho_\bq(\omega)\,
\rho_{-\bq}(\omega')\nonumber \\
&& \qquad\qquad \qquad \qquad \qquad \fract{\omega+\omega'}{\big(\omega+\omega'\big)^2 + \epsilon^2},\nonumber 
\eea
where the last expression is the leading order correction and $\rho_\bq(\omega)$ is the 
non-interacting density of states 
of particle-hole excitations at momentum transferred $\bq$. We note that the energy decreases as always in second order perturbation theory, which depends from our neglect of the Hartree term. 

On the other hand, let us consider the asymptotic $t\to\infty$ Hamiltonian $\mathcal{H}_* = 
\mathcal{H}_0 + U\,\mathcal{U}$, and the equilibrium internal energy at temperature $T\ll T_F$ ($\beta=1/T$ taking $K_B=1$), 
\bea
E(T) &=& - \fract{\partial}{\partial \beta}\,\ln\Tr\Big(\text{e}^{-\beta\mathcal{H}_*}\Big) 
\simeq E_0 + V\,\fract{\gamma_V}{2}\,T^2\label{E(T)}\\
&& 
-V\, \fract{U^2}{4}
\int \fract{d^3q}{(2\pi)^3}\,\int_0 d\omega\, d\omega'\,\, 
\fract{\rho_\bq(\omega)\,\rho_{-\bq}(\omega')}{\omega+\omega'},\nonumber
\eea
where $\gamma_V$ is the specific heat coefficient at constant volume, and we expanded at leading order 
in $T\sim U\ll T_F$.  According to the {\sl thermalization hypothesis}, the steady state averages of the system should coincide with equilibrium thermal averages at a temperature $T_*$ such that 
$E(T_*) = E_*$, which through Eqs. \eqn{1:Delta E} and \eqn{E(T)} leads, at leading order in $U$, 
to the expression 
\bea
T_*(U)^2 &\simeq& \fract{U^2}{2\gamma_V}
\int \fract{d^3q}{(2\pi)^3}\,\int_0 d\omega\, d\omega'\,\, 
\rho_\bq(\omega)\,
\rho_{-\bq}(\omega') \nonumber \\
&& \qquad  \fract{\epsilon^2}{\big(\omega+\omega'\big)\Big[\big(\omega+\omega'\big)^2 + \epsilon^2\Big]}\; + O(U^3)\nonumber\\
&\equiv& U^2\,g(\epsilon)^2 + O(U^3),\label{T_*} 
\eea
where $g(\epsilon) \sim \epsilon$ for small $\epsilon$ and tends to a constant for $\epsilon\to\infty$. 
It follows that $T_*$ starts linear in $U$,~\cite{Kollar} and vanishes in the adiabatic limit,  
$T_*\sim \epsilon\to 0$, as in this case the system must evolve into the ground state of the fully interacting Hamiltonian $\mathcal{H}_*$.  

\subsection{An operating assumption}

Let us then consider a generic observable $\mathcal{O}$ with time-dependent average $O(t,U)
= \bra{\Psi(t)}\mathcal{O}\ket{\Psi(t)}$. 
We shall denote as  
\be
O^{(m)}(t,U)= \fract{\partial^m O(t,U)}{\partial U^m},\label{O^m}
\ee
its Taylor coefficients.  
Similarly, 
\be
O_\text{eq}(T,U) = 
\fract{\Tr\Big(\text{e}^{-\beta\mathcal{H}_*}\;\mathcal{O}\Big)}
{\Tr\Big(\text{e}^{-\beta\mathcal{H}_*}\Big)},\label{O_eq}
\ee
defines the thermal average of $\mathcal{O}$ with the asymptotic Hamiltonian $\mathcal{H}_*$ 
at temperature $T=\beta^{-1}$.  
If thermalization occurs then 
\be
\lim_{t\to\infty}\,O(t,U) = O_\text{eq}\big(T_*(U),U\big),\label{therm-0}
\ee
which also implies that 
\be
\lim_{U\to 0}\lim_{t\to\infty}\, O^{(m)}(t,U) = \left(\fract{d^m O_{eq}\left(T_*(U),U\right)}{dU^m}\right)_{U=0}, \label{therm}
\ee
where the derivative on the r.h.s. acts both explicitly and implicitly through 
$T_*(U)$ of Eq. \eqn{T_*}. We observe that, while the r.h.s. of Eq. \eqn{therm}, which we shall hereafter assume finite for the observable under investigation, can be readily evaluated by equilibrium perturbation theory, the l.h.s. cannot unless it is possible to exchange the order of limits, i.e. 
\be
\lim_{U\to 0}\lim_{t\to\infty} O^{(m)}(t,U) \overset{?}{=}
 \lim_{t\to\infty} \lim_{U\to 0} O^{(m)}(t,U),\label{assumption}
\ee
since the r.h.s. is instead accessible by time-dependent perturbation theory. Should Eq. \eqn{assumption} be valid, we could evaluate independently both sides of Eq. \eqn{therm} and check if 
they coincide, hence whether thermalization holds. 

We first observe that for Eq. \eqn{assumption} 
to hold it is necessary that $O^{(m)}(t,0)$ does not grow indefinitely when $t\to\infty$. If the latter condition is fulfilled for any $m$, it is highly probable that Eq. \eqn{assumption} is valid, too. However, since it is practically unfeasible to check that $O^{(m)}(t\to\infty,0)$ is non-singular whatever $m$ is  
and, moreover, because we will be interested just in the second-order coefficient $m=2$, we shall 
assume that 
\begin{law}\label{1assumption}
If $\lim_{t\to\infty} O^{(m\leq 2)}(t,0)$ is finite, then Eq. \eqn{assumption} is valid at $m=2$.
\end{law}
This is not a rigorous statement as singularities for $m>2$ would 
invalidate the results also at $m=2$. As a matter of fact, such an assumption is implicitly taken in any 
perturbative calculation, since nobody can ever guarantee that higher order terms are non-singular making perturbation theory justified. In fact, Eq. \eqn{assumption} would be strictly valid should the time-dependent perturbation expansion be convergent or, if asymptotic, Borel summable,~\cite{Costin} which is not unlikely to be the case for the Dyson's series of an interacting Fermi gas. In addition, the exchange of the two limits, $t\to\infty$ and $U\to 0$, is expected to hold in the adiabatic limit, i.e. if we subsequently send $\epsilon\to 0$, which corresponds to the conventional equilibrium perturbation theory that is believed to be convergent for a three-dimensional Fermi liquid. Therefore, if the exchange of the two limits were not possible but for $\epsilon\to 0$, that should translate into a singular behaviour in $\epsilon$,  
which we do not find any trace of within time-dependent perturbation theory.

\subsection{A clever observable} 

In order to test the consistency of the thermalization hypothesis, we shall focus on the charge structure factor $S(\bq)$, i.e. the Fourier transform of the static density-density correlation function 
$S(\br) = \langle n_\br\,n_\bnot\rangle - n^2$, where $n_\br$ is the density operator and $n$ its average. 

At equilibrium, a three-dimensional Landau-Fermi liquid at zero temperature has equal-time density-density correlations that decay as a power-law 
\be
S(\br,U,T=0) \sim \fract{1}{r^4}.\label{S(r)-T=0}
\ee
However, at any finite temperature $T\not= 0$, 
\be
S(\br,U,T) \sim \exp\bigg(-\fract{2\pi T}{v_F}\,r\bigg),\label{S(r)-T}
\ee
decays exponentially, thus revealing the quantum-critical nature of the 
Fermi-liquid ground state that is spoiled as soon as temperature is turned on. 
We also mention that Eqs. \eqn{S(r)-T=0} and \eqn{S(r)-T} actually hold both in the 
absence and in the presence of the interaction $U$, which just affects the pre-factors and the Fermi velocity $v_F$ but not the exponent in \eqn{S(r)-T=0} that is universal and depends only on the dimensionality $d$, equal 3 in our case. In the language of critical phenomena, we could say that 
a repulsive interaction 
is a marginally irrelevant perturbation to the quantum-critical non-interacting Fermi sea, whereas temperature is relevant. 

Eqs. \eqn{S(r)-T=0} and \eqn{S(r)-T} can be readily derived in the absence of interaction. Here,   
\be
S_0(\br,T) =  n\,\delta_{\br\mathbf{0}} - 2G_0(\br,T)^2,\label{S_0-GG}
\ee
with  
$
G_0(\br,T)  = \langle c^\dagger_{\br\sigma} c^\dagga_{\mathbf{0}\sigma}\rangle_0
$, 
which, for $k_F\,r \gg 1$ reads 
\bea
G_0(\br,T)  &\simeq& 
- \frac{T k_F}{\pi v_F r\sinh\big( \pi T r/v_F\big)}\;\Big[ \cos \big(k_F r\big) \nonumber \\
&& - \fract{\pi T}{v_F k_F}\,\sin \big(k_F r\big)\, \coth \big(\pi T r/v_F\big)\Big],\label{G(r)}
\eea
being $k_F$ the Fermi momentum. $G_0(\br,T)$ thus decays as $r^{-2}$ at $T=0$ but exponentially 
at any $T\not = 0$. 
As a consequence, the Fourier transform $S_0(\bq,T)$ of  $S_0(\br,T)$ 
is non-analytic when $T=0$ at $|\bq|\to 0$ (as well as  
at $|\bq|\sim 2k_F$)
\be
S_0(\bq,T)=k_F^2\,|\bq|/6\pi^2,
\ee
but turns analytic as soon as $T\not = 0$, specifically 
\be
 S_0(\bq\to \bnot,T) = 2\,T\,\rho_0 + O(q^2),
\ee
 with $\rho_0$ the single-particle density of states at the chemical potential. For the reasons discussed above, the same change of analytic behavior occurs even for the interacting charge structure factor 
 $S(\bq,U,T)$; interaction just modifies the single-particle density of states turning it into the quasi-particle one. In conclusion, the analytic properties of the charge structure factor for $|\bq|\to  0$ neatly discriminate 
between power-law or exponentially decaying correlations in real space, see the Appendix \ref{App:sec1} for more details. 

Let us move back to our problem of interest. If thermalization indeed holds, according to the above discussion we must conclude that, while initially the structure factor $S(t,\bq,U)$, i.e. the Fourier transform 
of  $S(t,\br,U)=\bra{\Psi(t)} n_\br\,n_\bnot\ket{\Psi(t)} - n^2$, is equal to $S_0(\bq,0)$,  hence non-analytic, 
in the steady state  
\bea
\lim_{t\to\infty} S(t,\bq,U)  &\equiv& S_*(\bq,U) \label{therm:S(q)} \\
&=& S\big(\bq, U, T_*(U)\big) \sim 2\,T_*(U)\,\rho_0 + O(q^2), \nonumber 
\eea
namely its analytic properties should totally change during the time evolution. Within perturbation theory, 
such an analyticity switch 
must show up as a singularity in the perturbative expansion. Indeed, 
if we first expand $S(\bq,U,T)$ in $U$ and $T=T_*(U)\sim U$,
and only after we take $\bq\to\bnot$, then we would find through Eq. \eqn{T_*} that for small $|\bq|$ 
\bea 
&& \lim_{U\to 0}\,\fract{d^2 S\big(\bq,U,T_*(U)\big)}{dU^2} \simeq  
\fract{32\pi^2\rho_0^2}{k_F^2}\;\fract{g(\epsilon)^2}{|q|} \nonumber \\  
&&\qquad = \lim_{U\to 0}\,\lim_{t\to \infty} 
\fract{\partial^2 S(t,\bq,U)}{\partial U^2} 
.\label{therm.vs.notherm}
\eea
It follows from Eq. \eqn{therm.vs.notherm} that the second order Taylor coefficient of the steady-state structure factor must be singular as $|\bq|\to 0$ if 
thermalization occurs. This is what we shall try to verify under the Assumption \ref{1assumption} 
that specifically corresponds to approximate 
\bea
S_*(\bq,U) &\overset{?}{\simeq}& S_0(\bq) 
+ U\,  \lim_{t\to \infty}\,\lim_{U\to 0} \,
\fract{\partial S(t,\bq,U)}{\partial U}\nonumber \\
&& + \frac{1}{2}\,U^2\,  \lim_{t\to \infty}\,\lim_{U\to 0} \,
\fract{\partial^2 S(t,\bq,U)}{\partial U^2}, \label{2assumption} 
\eea
which we caution once more it is not rigorous unless the two limits $U\to 0$ and 
$t\to\infty$ commute.

\section{Second order perturbation theory results}
\label{Second order perturbation theory results}

The occurrence of thermalization can 
thus be readily confirmed by 
verifying whether the second order correction to $S(t\to\infty,U,\bq)$ is singular as $|\bq| \to 0$, 
a check that can be easily performed if Eq. \eqn{2assumption} is taken to be valid.   
This is an elementary though lengthy calculation, which we thoroughly describe in the Appendix 
\ref{App:sec2}. The outcome is however totally unexpected.  First of all, we find that the approach to the steady state is perfectly defined within perturbation theory and provided the thermodynamic limit is taken first. For instance, at first order and in the sudden quench limit, $\epsilon\to\infty$, we find that 
\bea
S(t,\bq,U) &\simeq&  S_0(\bq) \label{S(q,t)-final}\\ 
&& \!\!\!\! \!\!\!\!\!\!\!\!\!\!\!\! \!\! - 4U\!\int_0 d\omega_1\,d\omega_2\,
\rho_\bq(\omega_1)\,\rho_\bq(\omega_2)
\;\fract{1-\cos\left(\omega_1+\omega_2\right)t}{\omega_1
+\omega_2},\nonumber
\eea
where hereafter momentum and energy are in units of $k_F$ and the Fermi energy $\epsilon_F$, respectively. Since for small $|\bq|$ and 
 $\omega \leq v_F |\bq|$, the particle-hole density of states 
 $\rho_\bq(\omega)\sim \omega/v_F |\bq|$, 
the structure factor remains $\propto |\bq|$ at first order in $U$, and the time-dependent term in \eqn{S(q,t)-final} vanishes as a power-law in 
 $1/t$ for large times. In addition, and more remarkably, we find that in the steady-state the second order corrections are not singular as predicted by thermalization. Specifically and still for $\epsilon\to\infty$, 
 \be
S_*(\bq,U)  \sim A(U) + B(U)\,|\bq| + O(q^2),\label{end-S_*}
\ee
where 
\bea
A &=& 8\,U^2\,\int \fract{d\bp}{(2\pi)^3}\;
\iint_0 d\omega_1 d\omega_2\, \fract{\rho_\bp(\omega_1)\rho_\bp(\omega_2)}{\big(\omega_1+\omega_2\big)^2},
\label{delta-S-2nd-*-0}
\eea
and $B= 1/6\pi^2 + O(U)$, so  
that in real space $S_*(\br,U)$ still has a power-law decaying term that coexists with an exponentially vanishing one starting at order $U^2$. In other words, even though an exponential component arises as expected by thermalization, in contrast to it a power-law component persists in the steady state. 
We note that $S_*(\bq,U) \!\sim\! A + B\,|\bq| $ may also explains 
the contradictory one-dimensional results of Ref. ~\onlinecite{Alberto&Federico}.

We end by pointing out that the absence of singularities that we find up to second order, hence the apparent 
breakdown of thermalization, depends crucially 
on the vanishing particle-hole density of states at small frequency, $\rho_\bq(\omega)\sim \omega$, which 
is also the reason why Landau's adiabatic assumption is not unlikely in spite of the gapless spectrum. We believe this is not at all accidental.

\section{Out-of-equilibrium extension of Galitskii's low density approach}
\label{Bosonization} 

Going beyond leading order is in general unfeasible but in limiting cases where a consistent re-summation of the perturbative series is possible in terms
of expansion parameters different from interaction. At equilibrium this occurs for instance in a dense Fermi gas with long-range Coulomb forces, or, alternatively, in the dilute limit with
short-range interactions, where 
Galitskii~\cite{Galitskii} showed that replacing the full interaction vertex with the ladder diagrams in the particle-particle channel provides results valid at any order in interaction but at leading in the density. 
This was shown to actually correspond to diagonalizing the Hamiltonian in the subspace that includes, besides the Fermi sea, states with spin-singlet pairs of holes and particles, inside and  outside the Fermi sphere, respectively.~\cite{few-electrons} 
This approximate scheme is in turn close to Anderson's treatment of quantum  
fluctuation corrections to the BCS mean-field theory of superconductivity,~\cite{PWA-RPA} 
which we shall exploit to extend out-of-equilibrium Galitskii's theory. 
Indeed,  let us suppose that quantum fluctuations brought by interaction do not spoil completely the non-interacting Fermi sphere, which we shall regard as the {\sl vacuum} of 
the quantum fluctuations, hypothesis to be verified {\sl a posteriori}. We observe that 
\bea
&&\big[c^\dagger_{\bk\up}c^\dagger_{-\bk+\bqp\down},c^\dagga_{-\bp+\bq\down}c^\dagga_{\bp\up}\big]
= -\delta_{\bk\bp}\,c^\dagga_{-\bk+\bq\down}c^\dagger_{-\bk+\bqp\down} \nonumber \\
&&+ \delta_{-\bk+\bq\,,-\bp+\bqp}\,c^\dagger_{\bk\up}
c^\dagga_{\bk+\bqp-\bq\up},
\label{commutator}
\eea
has a finite value on the Fermi sea only if $\bk=\bp$ and $\bq=\bqp$, in which case is either +1 or -1 
depending whether $\bk$ and $-\bk+\bq$ are both inside or outside the Fermi sea, respectively. In the same spirit as e.g. bosonization, we approximate the r.h.s. of Eq. \eqn{commutator} by its average value on the Fermi sphere. Therefore, we associate to the pair creation operator 
$c^\dagger_{\bk\up}c^\dagger_{-\bk+\bq\down}$, where $\bk$ and $-\bk+\bq$ are both 
outside  the Fermi sphere, a hard-core boson creation operator $b^\dagger_{\bk,\bq}$. Seemingly, we associate another independent hard-core boson operator $a^\dagger_{\bk,\bq}$ to 
$c^\dagga_{-\bk+\bq\down}c^\dagga_{\bk\up}$, 
where now both $\bk$ and $-\bk+\bq$ are 
inside  the Fermi sphere. Since 
\be
\big[\mathcal{H}_0,c^\dagger_{\bk\up}c^\dagger_{-\bk+\bq\down}\big] \!=\! 
\big(\epsilon_\bk+\epsilon_{-\bk+\bq}\big)\, c^\dagger_{\bk\up}c^\dagger_{-\bk+\bq\down},
\ee
the non-interacting dynamics of the hard-core bosons can be reproduced by mapping
\be
\mathcal{H}_0 \rightarrow  \sum_\bq\,
\sum_{\bk}\,\omega_{\bk,\bq}\,\big(a^\dagger_{\bk,\bq}a^\dagga_{\bk,\bq} + 
b^\dagger_{\bk,\bq}b^\dagga_{\bk,\bq}\big), 
\ee
where $\omega_{\bk,\bq} = \left|\epsilon_{\bk}+\epsilon_{-\bk+\bq}\right| >0$. In this scheme, the Hamiltonian $\mathcal{H}=\mathcal{H}_0 + U\,\mathcal{U}$ is thus mapped onto~\cite{PWA-RPA}
\bea
\mathcal{H}_* &=& 
\sum_\bq\,\mathcal{H}_\bq = \sum_\bq\,
\bigg[
\sum_{\bk}\,\omega_{\bk,\bq}\,\big(a^\dagger_{\bk,\bq}a^\dagga_{\bk,\bq} + 
b^\dagger_{\bk,\bq}b^\dagga_{\bk,\bq}\big) \nonumber\\
&& + 
\fract{U}{V}\,\sum_{\bk,\bp}\,\big(b^\dagger_{\bk\bq} - a^\dagga_{\bk,\bq}\big)
\big(b^\dagga_{\bp\bq} - a^\dagger_{\bp,\bq}\big)
\bigg].\label{H-PWA}
\eea
The vacuum of the hard-core bosons is the Fermi sea, and the role of the interaction, second
term on the r.h.s. of Eq. \eqn{H-PWA}, is to create pairs of holes and particles out of the vacuum. 

We observe that each $\mathcal{H}_\bq$ in 
Eq. \eqn{H-PWA} resembles the Hamiltonian of hard-core bosons in the presence of a local potential. We thence  
foresee that $\langle a^\dagger_{\bk,\bq}a^\dagga_{\bk,\bq}\rangle\!\sim\! \langle b^\dagger_{\bk,\bq}b^\dagga_{\bk,\bq}\rangle\! \sim\! 1/V$. Consequently,  we expect it is safe to relax the hard-core constraint, and regard $a^\dagga_{\bk,\bq}$ and $b^\dagga_{\bk,\bq}$ as conventional bosons. 
Within this approximation, which we verified  {\sl a posteriori}, it is relatively straightforward to 
diagonalize Eq. \eqn{H-PWA}, see the Appendix \ref{App:sec3}.

The Hamiltonian \eqn{H-PWA} can be exploited to reproduce the dynamical behaviour 
of certain electronic observables following a sudden quench. For instance, the time-evolution 
of the momentum distribution, through the equation of motion 
$i\dot{n}_\bk = \big[\mathcal{H},n_\bk\big]$, maps onto 
\ba
n_\bp(t) \simeq 
 \begin{cases}
 1 - \sum_\bq\, \bra{\psi(t)} a^\dagger_{\bp,\bq}a^\dagga_{\bp,\bq}\ket{\psi(t)} &\text{if~} |\bp|\leq k_F,\\
 \sum_\bq\, \bra{\psi(t)} b^\dagger_{\bp,\bq}b^\dagga_{\bp,\bq}\ket{\psi(t)}&\text{if~} |\bp|> k_F, 
 \end{cases}\label{n_k}
\ea
so that the jump at the Fermi surface
\be
Z(t) \simeq 1 - \sum_\bq\, \bra{\psi(t)} a^\dagger_{\bk,\bq}a^\dagga_{\bk,\bq}
+ b^\dagger_{\bk,\bq}b^\dagga_{\bk,\bq}\ket{\psi(t)} 
,\label{Z-expression}
\ee
where $|\bk|$ is on the Fermi sphere and $\ket{\psi(t)}$ is the boson-vacuum evolved with the Hamiltonian \eqn{H-PWA}. 
Through the exact diagonalization of the latter,  
assuming a spherical Fermi surface with energy dispersion 
$\epsilon_\bk = k^2$, which is indeed appropriate in the low-density limit, we obtain a steady state value of the jump at the Fermi surface $Z_*=Z_{eq}+\delta Z_*$, see the Appendix \ref{App:sec3}, where 
\bw
\bea
\!\!\!\!\!\!\delta Z_* \!\! &=& \!\! - 
\fract{(k_F a)^3}{8\pi^2}\!\!\int_0^2 \!\!\!q\,dq\,
\Bigg[\! \int_0^{q(2-q)}\!\!\!\!\!\!\!\!d\omega\,
\Big|T\big( \! -\omega+i0^+,\bq\big)\Big|^2\! \int_0\!d\ep\, 
\fract{\mathcal{N}_\text{OUT}(\ep,\bq)}{\big(\omega +\ep\big)^2}
+ \! \int_0^{q(2+q)}\!\!\!\!\!\!\!\!d\omega\,
\Big|T\big(\omega-i0^+,\bq\big)\Big|^2\!
\int_0\!d\ep\, 
\fract{\mathcal{N}_\text{IN}(\ep,\bq)}{\big(\omega+\ep\big)^2}\,
\Bigg]\!,\;\label{Z-off-eq}
\eea
 with $a$ the lattice spacing and $Z_{eq}$  
the equilibrium value at zero temperature~\cite{Galitskii}
\ba
Z_\text{eq.} \! &=& \! 1 - 
\fract{(k_F a)^3}{8\pi^2}\int_0^2 q\,dq
\Bigg[ \int_0^{q(2-q)}\!\!\!\! \!\! \! \!\!d\omega\,
\int_0 \fract{d\ep}{\pi}\, 
\fract{\Im\text{m} T\big(\ep-i0^+,\bq\big)}{\big(\ep+\omega\big)^2}
+ \int_0^{q(2+q)}\!\!\!\! \!\! \! \!\!d\omega\,
 \int_0 \fract{d\ep}{\pi}\,
 \fract{\Im\text{m} T\big( \! -\ep+i0^+,\bq\big)}{\big(\ep+\omega\big)^2}\,
\Bigg].
\ea
\ew

The function of complex variable $T(z)$ 
defined through
\bea
U\,T^{-1}(z,\bq) &=& 1 - U\,\chi(z,\bq) 
= 1 + 
U\,\int_0\,d\ep\;\fract{\mathcal{N}_\text{IN}(\ep,\bq)}{z+\ep}\nonumber\\
&& - U\,\int_0\,d\ep\;\fract{\mathcal{N}_\text{OUT}(\ep,\bq)}{z-\ep},
\label{def:T-matrix}
\eea
is just the usual $T$-matrix, with $\chi(z,\bq)$ the non-interacting Cooper bubble,  
$\mathcal{N}_\text{IN}(\ep,\bq)$ and $\mathcal{N}_\text{OUT}(\ep,\bq)$ the density of states of a pair of holes and 
particles, respectively, at total momentum $\bq$. In particular, if we expand up to second order in $U$ we find that 
$
1-Z_* = 2\big(1-Z_\text{eq.}\big),
$
in agreement with Ref.~\onlinecite{Kehrein}. 

Following Galitskii~\cite{Galitskii}, 
if $\chi_0(z,\bq)$ is the Cooper bubble of two electrons in the vacuum, 
then 
\be
T(z,\bq)  \simeq T_0(z,\bq) + T_0(z,\bq)^2\,
\Big(\chi(z,\bq)-\chi_0(z,\bq)\Big),\label{T-expanded}
\ee
where $T_0(z,\bq)$
is the scattering $T$-matrix in the vacuum. 
Using the first order expansion \eqn{T-expanded} in Eq. 
\eqn{Z-off-eq}, one consistently obtains a value that is exact at any order in the interaction $U$, 
but valid up to second order in $T_0(z\to 0)\sim k_F f_0$, where $f_0$ is the scattering length of two particles in the vacuum~\cite{Galitskii}.

We thus find that $Z_*$ in the steady state remains 
strictly finite, though smaller than at equilibrium, at leading order in the density but infinite in $U$. The simplest interpretation of this result is that  
the momentum distribution jump does not thermalize at low density at any order in perturbation theory, which is consistent with the previous second order calculation. Indeed the discontinuity of 
$n_\bk$ at the Fermi surface implies a Friedel-like behaviour of the steady-state single-particle density matrix $G_*(\br) = 
\langle c^\dagger_{\br\sigma}c^\dagga_{\mathbf{0}\sigma}\rangle_{t\to\infty}\sim \cos\big(k_F r\big)/r^2$, at odds with the thermalization prediction of an exponential decay. 

Even though the above results are valid at any order in $U$, still an expansion in the small parameter 
$k_F\,f_0$ is assumed. Therefore also in this case in order for the results to be representative of the actual 
steady-state we have to assume like before that the two limits of $t\to\infty$ and $k_F f_0\to 0$ commute 
for the Taylor coefficients of the expansion in powers of $k_F f_0$. 

\section{Discussion} 
\label{Conclusions}

We have presented two separate calculations both of which show that the power-law decay of the equal-time correlations characteristic of a Fermi sea seems to survive the switching of a very weak interaction, even when not adiabatically slow. This contrasts the expectation that the excess energy supplied in the switching process, when it is not adiabatic, should heat the system hence effectively rise its internal temperature. The common-sense reaction to those results would be they are not valid because time-dependent perturbation theory is unjustified as $t\to\infty$. The argument would be that, even if we have not encountered any singularity as the time $t\to\infty$ up to second order, it is certain that at some higher order a singularity  arises invalidating also the second order calculation. We emphasise that should perturbation theory be indeed 
ill-defined, then a singularity should be accessible in perturbation theory simply because 
the effective temperature that corresponds to the injected energy is perturbative in the interaction. Therefore, 
if an exponential decay $\text{e}^{-\gamma\,t}$ arises, which we do not find evidence for up to second order, the decay rate must be perturbative in the interaction, hence at some high order it must show up through a correction growing linearly with $t$. Evidently we cannot exclude this is indeed the case, hence that our second order calculation does not allow to conclude that thermalization is absent. 

Even though a lot of reasonable arguments can be invoked to argue that thermalization must finally take place, nevertheless there are also counterarguments that one can envisage, some of which we shall list here, which leave open this issue hence worth to be further investigated.  

\subsection{Fermi liquid many-body spectrum}

According to Landau,\cite{Landau-1,Landau-2} a low excitation energy $\delta E$ of a Fermi liquid can be parametrized in terms of the deviations $\delta n_{\bk\sigma}$ from equilibrium of the quasiparticle occupation numbers at momentum $\bk$ and spin $\sigma$,  and Taylor expanded as 
\bea
\delta E\big[\delta n\big] &\simeq& \sum_{\bk\sigma}\,\epsilon_{\bk*}\,\delta n_{\bk\sigma}
\label{Landau:deltaE}\\
&& + \fract{1}{2}\,\sum_{\bk\bkp,\sigma\sigma'}\,
f_{\bk\sigma,\bkp\sigma'}\, \delta n_{\bk\sigma}\,\delta n_{\bkp\sigma'} + O\Big(\delta n^3\Big). 
\nonumber
\eea
Because of the correspondence between non-interacting and interacting low-energy many-body eigenstates, 
the entropy of a non-equilibrium quasiparticle population coincides with that of non-interacting particles, which, together with Eq.\eqn{Landau:deltaE}, allow to calculate finite temperature properties that it is well known reproduce the correct thermodynamic 
behaviour of a Fermi liquid for temperature $T\lesssim T_F$. 
This implies that Eq.\eqn{Landau:deltaE} describes the many-body eigenvalue spectrum over a macroscopic energy interval above the ground state, $\Delta E \lesssim V\,T_F$. Unsurprisingly, the level statistics that corresponds to the energy functional \eqn{Landau:deltaE} has a Poisson distribution~\cite{regis1,regis2} 
characteristic of an integrable system, even though a three-dimensional interacting Fermi gas is supposedly not integrable. It was suggested in Ref. \onlinecite{distasio1} that a non-integrable model whose low-energy properties can be described by an integrable effective Hamiltonian, 
which is the case of a Landau-Fermi liquid, will posses Poisson's level statistics of the low-energy many-body spectrum, while the 
full many-body spectrum will obviously have a Wigner-Dyson's distribution. If that conjecture were true, it would imply no equipartition of the low energy density 
$\ll T_F$ that is injected during the switching process, which might explain our results. 

\subsection{Locality of observables and thermalization}

Rigorously speaking, thermalization can be justified only when the Hamiltonian density is local and in reference
to local observables. In this situation, one reasonably expects that the reduced density matrix of a subsystem, with the rest of the system playing the role of a dissipative bath,    
will evolve towards a Maxwell-Boltzmann distribution in the steady state, so that any subsystem observable will indeed thermalize. 

On the contrary, long range correlations are non-local observables hence it is not guaranteed they are going to thermalize. Indeed, while the interaction 
\be
U\,\mathcal{U} = U\,\sum_\bR n_{\bR\up}\,n_{\bR\down},
\ee
is local in real space, it is infinitely long-ranged in momentum space 
\be
U\, \mathcal{U} = \fract{U}{V}\,\sum_{\bk,\bp,\bq} 
c^\dagger_{\bk\up}c^\dagger_{\bp+\bq\down}c^\dagga_{\bp\down}c^\dagga_{\bk+\bq\up}.
\ee
Therefore it is not evident that an observable local in momentum space may thermalize. 

Note that the restriction of the notion of thermalization to observables that are local in real space would have no contradiction with our findings and no inconsistency as well with the adiabatic assumption at the basis of Landau's Fermi-liquid theory, since the latter  mainly concerns long-wavelength properties.  Indeed, the long-wavelength limit of the static structure factor or the momentum distribution close to $k_F$ are, rigorously speaking, not ``local observables".

\subsection{Relevance/irrelevance of interaction at the Fermi liquid quantum critical point}

We mentioned several times that a Fermi sea can be regarded as a quantum critical state of matter. At zero temperature all equal-time correlations decay as a power-law in the distance, with universal exponents that depend only on the dimensionality $d$, while at any finite temperature the decay turns exponential in the distance. 
From the viewpoint of criticality, a repulsive interaction in $d>1$ 
plays the role of a marginally irrelevant perturbation that only changes the pre-factors of correlation functions but does not affect the exponent of their power-law decay. 

In the language of critical phenomena one could be tempted to conclude that the non-interacting Fermi sea critical state is stable towards switching a weak repulsion or, more generally, that the Landau-Fermi liquid critical state survives weak changes of the repulsion  strength. If thermalization occurred, that statement would be incorrect, since no matter how weak and slow the change of interaction is, a finite temperature will be eventually generated driving the system away from criticality. 

The above question whether the irrelevant character of a perturbation with respect to a quantum critical point remains unaltered also when the switching process is not adiabatic is to our knowledge still open.~\cite{dallatorre,Mitra2011} Therefore we cannot exclude that the power-law decay of correlations in a Fermi sea does survive the turning on of a weak repulsion, hence that our results are correct.

\begin{acknowledgments}
We acknowledge useful discussions with F. Becca, A. Parola and S. 
Cecotti.  
This work has been supported by the European Union, Seventh Framework Programme, under the project GO FAST, Grant Agreement no. 280555. 
\end{acknowledgments}



\appendix

\section{Remarks on the long-wavelength structure factor $S(\bq)$}
\label{App:sec1}

Here we recall the connection between the analytic properties of the long-wavelength structure factor $S(\bq)$ and the behavior in real space of its inverse Fourier transform, $S(\br)$. In three dimensions and assuming space isotropy $S(\br)=S(r)$, we find   
\bea
S(\bq) &=& S(q) = \int d\br\,S(\br)\,\text{e}^{-i\bq\cdot\br} \nonumber\\
&=& \fract{4\pi}{q}\;\int_0^\infty r dr\, S(r)\,
\sin(q r).\label{SM-S(q)}
\eea
Even though $q=|q|$ is by definition positive, if we extend artificially the domain of $S(q)$ also to $q\leq 0$, we readily note that $S(q)=S(-q)$ in an even function. Therefore the Taylor expansion of $S(q)$ near  
$q=0$ must contain even powers, $q^{2n}$, and/or odd powers but of the absolute value, $|q|^{2n+1}$.  In the  latter case the function $S(q)$, $q\in[-\infty,+\infty]$, is not analytic at the origin (and presumably also 
at $q=\pm 2k_F$). 

Through Eq. \eqn{SM-S(q)}, we observe that, if $S(r)$ vanishes faster than any power of $1/r$ 
for $r\to\infty$, then only even-order derivatives of $S(q)$ are finite for $q\to 0$; the function is analytic. On the contrary, if $S(r)\sim 1/r^{2m}$ for large $r$ and $m\geq 2$, then the derivative of order $2m-3$ will be finite at $q=0$,  
i.e. $S(q)$ will have a non-analytic behavior 
$S(q) \sim |q|^{2m-3}$ at small $q$. In particular, 
if 
\[
S(r\to\infty) \sim A\,\text{e}^{-r/\xi} + \fract{B}{r^4},
\]
then 
\[
S(q\to 0) \sim 8\pi\,\xi^3\,A - \pi^2\,B\,|q| + O(q^2).
\]
Therefore, even though the term linear in $q$ is sub-leading 
with respect to the constant, its inverse Fourier transform corresponds to a power-law decaying contribution, which dominates over the exponentially vanishing one that derives from the leading constant-in-$q$ term $8\pi\,\xi^3\,A$. In other words, what discriminates between a power-law with respect to an exponential decay of $S(r)$ for large $r$ is just the finiteness of odd-order derivatives of $S(q)$ at $q=0$.   
  
\section{Second order perturbation theory}
\label{App:sec2}

We consider a Hamiltonian
\be
\mathcal{H} = \mathcal{H}_0 + U(t)\, \mathcal{U},\label{Ham}
\ee
where the unperturbed  $\mathcal{H}_0$  has eigenstates $\ket{n}$ 
with eigenvalues $E_n$, measured with respect to the ground state energy.  We shall assume that 
$\bra{n}\mathcal{U}\ket{n}= 0$ for any $\ket{n}$ and, in addition, that the Hamiltonian as well as the eigenvalues are real. 

We shall here consider a very general turning on of the interaction,  
\be
U(t) = U\,\left(1-\text{e}^{-\epsilon t}\right),
\ee
with $\epsilon>0$, which interpolates between the sudden switch, $\epsilon\to\infty$, and the adiabatic 
one, send first $t\to\infty$ and only later $\epsilon\to 0$. The reason is that, as we shall see, there is not a dramatic difference between the adiabatic $\epsilon\to 0$ limit and the general case of finite $\epsilon$, at least up to second order. Since the former is known not to lead to singularities as 
$t\to \infty$, this suggests the same holds for any $\epsilon\not = 0$. 

We assume that the system is initially in the ground state $\ket{0}$ of $\mathcal{H}_0$, and evolves 
at positive times with the interacting Hamiltonian \eqn{Ham}. We study the time evolution 
by second order perturbation theory applied directly to the Schr{\oe}dinger equation in the interaction representation. Namely we write the wavefunction as ($\hbar=1$) 
\be
\ket{\Psi(t)} =  \text{e}^{-i\mathcal{H}_0 t}\; \ket{\Phi(t)},\label{initial-Psi}
\ee
where $\ket{\Phi(0)}=\ket{\! 0}$, and set $\ket{\Phi(t)}=\sum_{m\geq 0}\, U^m\,\ket{\phi_m(t)}$, where $m$ is the order in perturbation theory, being $\ket{\phi_0(t)}= \ket{0}$ and, for any $m>0$, $\ket{\phi_m(0)}=0$. It readily follows that 
\ba
i\,\partial_t \ket{\phi_m(t)} = \frac{U(t)}{U}\; \text{e}^{i\mathcal{H}_0 t}\, 
\mathcal{U}\,\text{e}^{-i\mathcal{H}_0 t}\ket{\phi_{m-1}(t)},
\ea
which leads to 
\bw
\bea
\text{e}^{-i\mathcal{H}_0 t}\;\ket{\phi_1(t)} &=&
\sum_{n\not= 0}\, \Bigg( \fract{\text{e}^{-i E_n t}-1}{E_n}
- \fract{\text{e}^{-i E_n t}-\text{e}^{-\epsilon t}}{E_n +i\epsilon}\Bigg)\,W_{n0}\,
\ket{n} 
\equiv \sum_{n\not = 0}\,W_{n0}\,
\Lambda_\epsilon(E_n;t)\ket{n},\label{phi_1}\\
\text{e}^{-i\mathcal{H}_0 t}\;\ket{\phi_2(t)} &=&  
\Bigg\{  \sum_{n\not = 0}\, \left|W_{n0}\right|^2 \bigg[
\fract{1}{E_n^2+\epsilon^2} - \fract{1}{E_n^2}
+ \fract{\text{e}^{-iE_n t}}{E_n}\;\Big(\frac{1}{E_n}-\frac{1}{E_n+i\epsilon}\Big)
- \fract{\text{e}^{-i\left(E_n-i\epsilon\right) t}}{E_n-i\epsilon}\;\Big(\frac{1}{E_n}-\frac{1}{E_n+i\epsilon}\Big)\nonumber \\
&& +i\,\fract{1}{E_n}\;\left(t - \fract{1-\text{e}^{-\epsilon t}}{\epsilon}\right)
-i \,\fract{1}{E_n+i\epsilon}\; \fract{\left(1-\text{e}^{-\epsilon t}\right)^2}{2\epsilon} 
\bigg] \Bigg\}\ket{0}\nonumber\\
&& +\sum_{n\not = 0}\,\ket{n}\;\sum_{m\not= 0,n}\,W_{nm}\,W_{m0}\,\Bigg\{
\frac{1}{E_m}\,\bigg[\fract{\text{e}^{-i E_n t}-\text{e}^{-i E_m t}}{E_n-E_m} 
- \fract{\text{e}^{-i E_n t}-\text{e}^{-i E_m t-\epsilon t}}{E_n-E_m+ i\epsilon}\bigg]\nonumber\\
&& - \frac{1}{E_m}\,\bigg[\fract{\text{e}^{-i E_n t}-1}{E_n} 
- \fract{\text{e}^{-i E_n t}-\text{e}^{-\epsilon t}}{E_n+ i\epsilon}\bigg]\nonumber\\
&& -\frac{1}{E_m+i\epsilon}\,\bigg[\fract{\text{e}^{-i E_n t}
-\text{e}^{-i E_m t}}{E_n-E_m} 
- \fract{\text{e}^{-i E_n t}
-\text{e}^{-i E_m t-\epsilon t}}{E_n-E_m+ i\epsilon}\bigg]
+\frac{1}{E_m+i\epsilon}\,\bigg[\fract{\text{e}^{-i E_n t}
-\text{e}^{-\epsilon t }}{E_n+i\epsilon} 
- \fract{\text{e}^{-i E_n t}
-\text{e}^{-2\epsilon t}}{E_n+ 2i\epsilon}\bigg]\Bigg\}
\nonumber\\
&\equiv& \Big(A(t)-1\Big) \ket{0} + \sum_{m\not = 0,n}\,
\sum_{n\not = 0}\, W_{mn}\,W_{n0}\;\Xi_\epsilon(E_m,E_n;t)\ket{m},
\label{phi_2}
\eea
\ew
with $W_{nm}=\bra{n} \mathcal{U}\ket{m}$. 

In the specific case of the Hubbard interaction,
\bea
\mathcal{H} &=& \sum_{\bk\sigma}\,\epsilon_\bk\,
c^\dagger_{\bk\sigma}c^\dagga_{\bk\sigma}
+ \frac{U(t)}{V}\,\sum_{\bk\bp}\,\sum_{\bQ\not=\bnot}\,
c^\dagger_{\bk\up}c^\dagger_{\bp+\bQ\down}c^\dagga_{\bp\down}
c^\dagga_{\bk+\bQ\up}\nonumber \\
&=& \mathcal{H}_0 + U(t)\,\mathcal{U},\label{SM-Hubbard}
\eea
where the $\bQ=\bnot$ term is not included so to fulfill $W_{nn}=0$ 
, $\forall \ket{n}$, the matrix element $W_{n0}$, 
where $\ket{0}$ is the unperturbed Fermi sea, is finite 
only if 
\bea
\ket{n} &=& c^\dagger_{\bk\up}c^\dagger_{\bp+\bQ\down}
c^\dagga_{\bp\down}c^\dagga_{\bk+\bQ\up}\ket{0},\label{ket-n}
\eea
where $|\bk|>k_F$ and $|\bp+\bQ|>k_F$, hence refer to particles, while 
$|\bk+\bQ|\leq k_F$ and $|\bp|\leq k_F$, hence refer to holes. The energy of this state is 
$E_n=\omega_{\bk,\bk+\bQ} + \omega_{\bp+\bQ,\bp}= 
\big(\epsilon_\bk-\epsilon_{\bk+\bQ}\big) 
+\big(\epsilon_{\bp+\bQ}-\epsilon_{\bp}\big) >0$, 
where energies are measured with respect to the chemical potential. The matrix element is $W_{n0}=1/V$. At second order we must account also for states with three and four particle-hole pairs besides those with two.  

We write the wavefunction up to second order as 
\be
\ket{\Psi(t)} = A(t)\, \ket{0} + 
\ket{\psi_1(t)}  
+ \ket{\psi_{2}(t)},
\ee
where $A(t)=1+O(U^2)$, $\bra{0}\psi_{1}(t)\rangle=0=\bra{0}\psi_{2}(t)\rangle$ and 
\be
\mid A(t)\mid^2 + \bra{\psi_1(t)}\psi_1(t)\rangle = 1 + O(U^3).
\label{normalization}
\ee 
We are interested in calculating up to second order the 
average of the structure-factor operator 
$\mathcal{S}_\bq = n_\bq n_{-\bq}/V$ at $\bq\not = \mathbf{0}$, which reads, through 
Eqs. \eqn{normalization}, \eqn{phi_1} and \eqn{phi_2},
\bw
\bea
S(\bq,t) &\simeq& S_0(\bq) 
+ \Big(\bra{\psi_1(t)} \mathcal{S}_\bq\ket{0} + c.c.\Big)
+ \Big(\bra{\psi_2(t)} \mathcal{S}_\bq\ket{0} + c.c.\Big)
+ \bra{\psi_1(t)} \mathcal{S}_\bq-S_0(\bq)\ket{\psi_1(t)} \nonumber\\
&=& S_0(\bq) + U\,\sum_{n\not = 0} \bigg[W_{0n}\,\Lambda_\epsilon(E_n,t)^*\,
\bra{n} \mathcal{S}_\bq \ket{0} + c.c.\bigg] + U^2\,\sum_{n\not = 0,m}\sum_{m\not = 0}\,
\bigg[W_{0m}W_{mn}\,\Xi_\epsilon(E_n,E_m;t)^* \bra{n} \mathcal{S}_\bq \ket{0}+ c.c.\bigg] 
\nonumber\\
&& + U^2\,\sum_{n,m\not = 0}\, W_{0m}\,W_{n0}\,\Lambda_\epsilon(E_m;t)^*\,\Lambda_\epsilon(E_n;t)\; \bra{m} \mathcal{S}_\bq -S_0(\bq)\ket{n}
\label{2nd-order-Sq} 
\eea
\ew
where $S_0(\bq) = \bra{0} \mathcal{S}_\bq\ket{0}$ is the structure factor of the Fermi sea. 

The equilibrium perturbation theory can be recovered 
in the adiabatic limit, which amounts 
in Eqs. \eqn{phi_1} and \eqn{phi_2} to set first $\epsilon t \to\infty$ and then send $\epsilon\to 0$. Indeed, in this limit we obtain 
\bea
\Lambda_{\epsilon\to 0}(E_n;t) &=& -\fract{1}{E_n},
\label{adia-1}\\
\Xi_{\epsilon\to 0}(E_n,E_m;t) &=& 
\frac{1}{E_m\,E_n},\label{adia-2}
\eea 
which are the well known expansion coefficients at second order and at equilibrium. Therefore the equilibrium perturbative expansion of the structure factor is 
\bw
\bea
S_\text{eq}(\bq) &=& \lim_{\epsilon\to 0}  
S(\bq,t\gg 1/\epsilon) = 
S_0(\bq)  
 - U\,\sum_{n\not = 0} \bigg[\fract{W_{0n}}{E_n}\;
\bra{n} \mathcal{S}_\bq \ket{0} + c.c.\bigg] + U^2\,\sum_{n\not = 0,m}\sum_{m\not = 0}\,
\bigg[\fract{W_{0m}W_{mn}}{E_n E_m}\, \bra{n} \mathcal{S}_\bq \ket{0}+ c.c.\bigg] 
\nonumber\\
&& + U^2\,\sum_{n,m\not = 0}\, \fract{W_{0m}\,W_{n0}}{E_n E_m}\; \bra{m} \mathcal{S}_\bq -S_0(\bq)\ket{n},\label{Sq-equilibrium}
\eea 
\ew
and we know it does not contain any singular term. 

However, we are actually interested in the opposite limit of a sudden switch, which amounts 
to send $\epsilon\to\infty$ so that 
\bea
\Lambda_{\epsilon\to\infty}(E_n;t) &=& -\fract{1-\text{e}^{-iE_n t}}{E_n},\label{sudden-1}\\
\Xi_{\epsilon\to\infty}(E_n,E_m;t) &=& 
\frac{1}{E_m}\;\fract{\text{e}^{-i E_n t}-\text{e}^{-i E_m t}}{E_n-E_m} 
\nonumber\\
&& - \frac{1}{E_m}\;\fract{\text{e}^{-i E_n t}-1}{E_n}.
\label{sudden-2}
\eea 
By means of Eqs. \eqn{2nd-order-Sq} and \eqn{Sq-equilibrium}, and observing that all matrix elements are real, we can write the structure factor in the sudden limit as
\bw
\begin{subequations}
\begin{align} 
S(\bq,t) =& S_{\text{eq.}}(\bq) + U\,\sum_{n\not = 0} \bigg[W_{0n}\,
\Big(\Lambda_{\epsilon\to\infty}(E_n,t)^* - \Lambda_0(E_n,t)^*\Big)
\bra{n} \mathcal{S}_\bq \ket{0} + c.c.\bigg] \nonumber \\
& + U^2\,\sum_{n\not = 0,m}\sum_{m\not = 0}\,
\bigg[W_{0m}W_{mn}\,\Big(\Xi_{\epsilon\to\infty}(E_m,E_n;t)^*
-\Xi_0(E_m,E_n;t)^*\Big)
 \bra{n} \mathcal{S}_\bq \ket{0}+ c.c.\bigg] 
\nonumber\\
& + U^2\,\sum_{n,m\not = 0}\, W_{0m}\,W_{n0}\,
\Big(\Lambda_{\epsilon\to\infty}(E_m;t)^*\,
\Lambda_{\epsilon\to\infty}(E_n;t)
- \Lambda_0(E_m;t)^*\,\Lambda_0(E_n;t)\Big)
\; \bra{m} \mathcal{S}_\bq - S_0(\bq)\ket{n}\nonumber\\
=& S_{\text{eq.}}(\bq) + 2U\,\sum_{n\not = 0} W_{0n}\,
\fract{\cos( E_n t)}{E_n}\;
\bra{n} \mathcal{S}_\bq \ket{0}  \label{2nd-order-Sq-zero} \\
& + 2U^2\,\sum_{n\not = 0,m}\sum_{m\not = 0}\,
W_{0m}W_{mn}\,\bigg(
\frac{1}{E_m}\;\fract{\cos(E_n t)-\cos(E_m t)}{E_n-E_m} 
 - \frac{1}{E_m}\;\fract{\cos(E_n t)}{E_n}\bigg)
 \bra{n} \mathcal{S}_\bq \ket{0} 
\label{2nd-order-Sq-uno}\\
& + U^2\,\sum_{n,m\not = 0}\, W_{0m}\,W_{n0}\,
\fract{\cos(E_m-E_n)t -\cos( E_m t)-\cos(E_nt)}{E_n E_m}
\; \bra{m} \mathcal{S}_\bq -S_0(\bq)\ket{n}.
\label{2nd-order-Sq-bis} 
\end{align}
\end{subequations}
\ew
In the thermodynamic limit the 
second term in Eq. \eqn{2nd-order-Sq-zero} can be written as 
\bea
\delta S^{(1)}(\bq,t) &=& 4U\,\iint_0 d\omega_1\,d\omega_2 \,
\rho_\bq(\omega_1)\,\rho_\bq(\omega_2)
\nonumber\\
&& \qquad\qquad \fract{\cos\big(\omega_1+\omega_2\big)t}{\omega_1+\omega_2},
\label{delta-S-1st}
\eea
where $\rho_\bq(\omega)$ is the density of states of a particle-hole
excitation at momentum transferred $\bq$. The denominator vanishes  at small frequencies but this singularity is canceled by the 
numerator since, 
for $\omega \ll v_F q$, $\rho_\bq(\omega)\sim \theta(2k_F-q)\,\omega/v_F q$. As a result, in the long time limit  
Eq. \eqn{delta-S-1st} vanishes with a power law in $1/t$, so that the steady-state $S_*(\bq)$ after a sudden quench  
coincides with the equilibrium $S_\text{eq.}(\bq)$ up to first order.  

The explicit evaluation of Eqs. \eqn{2nd-order-Sq-uno} and 
\eqn{2nd-order-Sq-bis} as well as their final expressions are 
simple though quite lengthy. Therefore we prefer to 
show graphically all terms that contribute. Their matrix elements  have all the same 
absolute value, equal to $1/V^3$, apart from a sign that is indicated in the figures. 

\begin{figure}[hbt]
\centerline{\includegraphics[width=7cm]{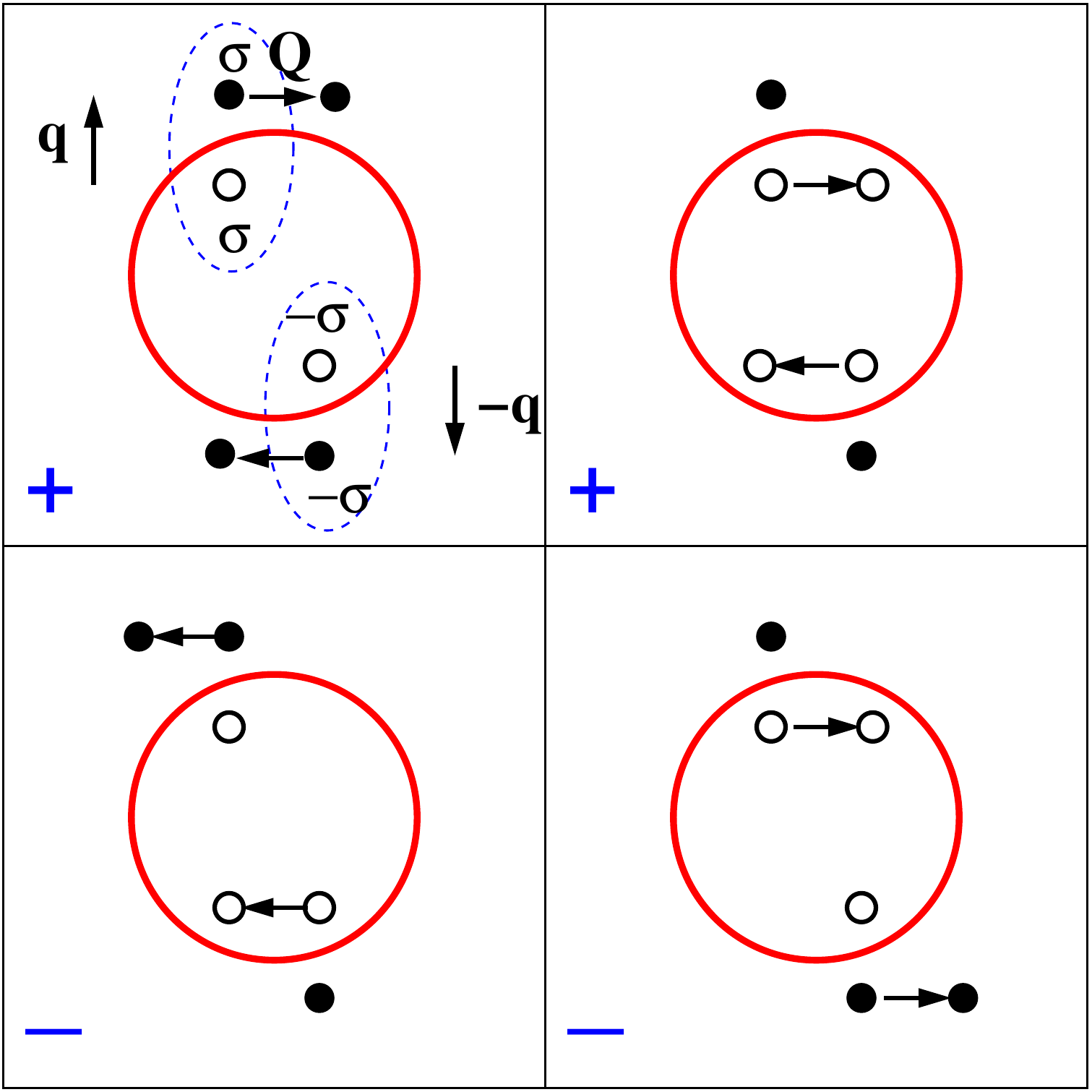}}
\caption{(Color online) Scattering processes that refer to Eq. \eqn{2nd-order-Sq-uno}. The circle represents the Fermi sphere, open dots holes and solid ones particles. The initial particle-hole pairs in $\ket{n}$ are encircled and have opposite spins. The scattering process, i.e. the motion of particles/holes indicated by the arrows, leads to a new state $\ket{m}$. }
\label{fig-1}
\end{figure}
Let us first consider the term in Eq. \eqn{2nd-order-Sq-uno}. 
The state $\ket{n}$ includes two particle-hole (p-h) pairs at momentum transferred $\bq$ and $-\bq$ and spin $\sigma$ and $\sigma'$. 
The intermediate state $\ket{m}$ can be reached by the interaction both from $\ket{n}$ and from $\ket{0}$, hence it contains two 
p-h pairs with opposite spin and transferred momenta. 
If $\sigma'=-\sigma$, 
the processes that bring $\ket{n}$ to $\ket{m}$ are shown in 
Fig. \ref{fig-1}. In the figure $\ket{m}$ contains two opposite-spin 
p-h pairs at momenta $\bq+\bQ$ and $-\bq-\bQ$.

\begin{figure}[t]
\centerline{\includegraphics[width=7cm]{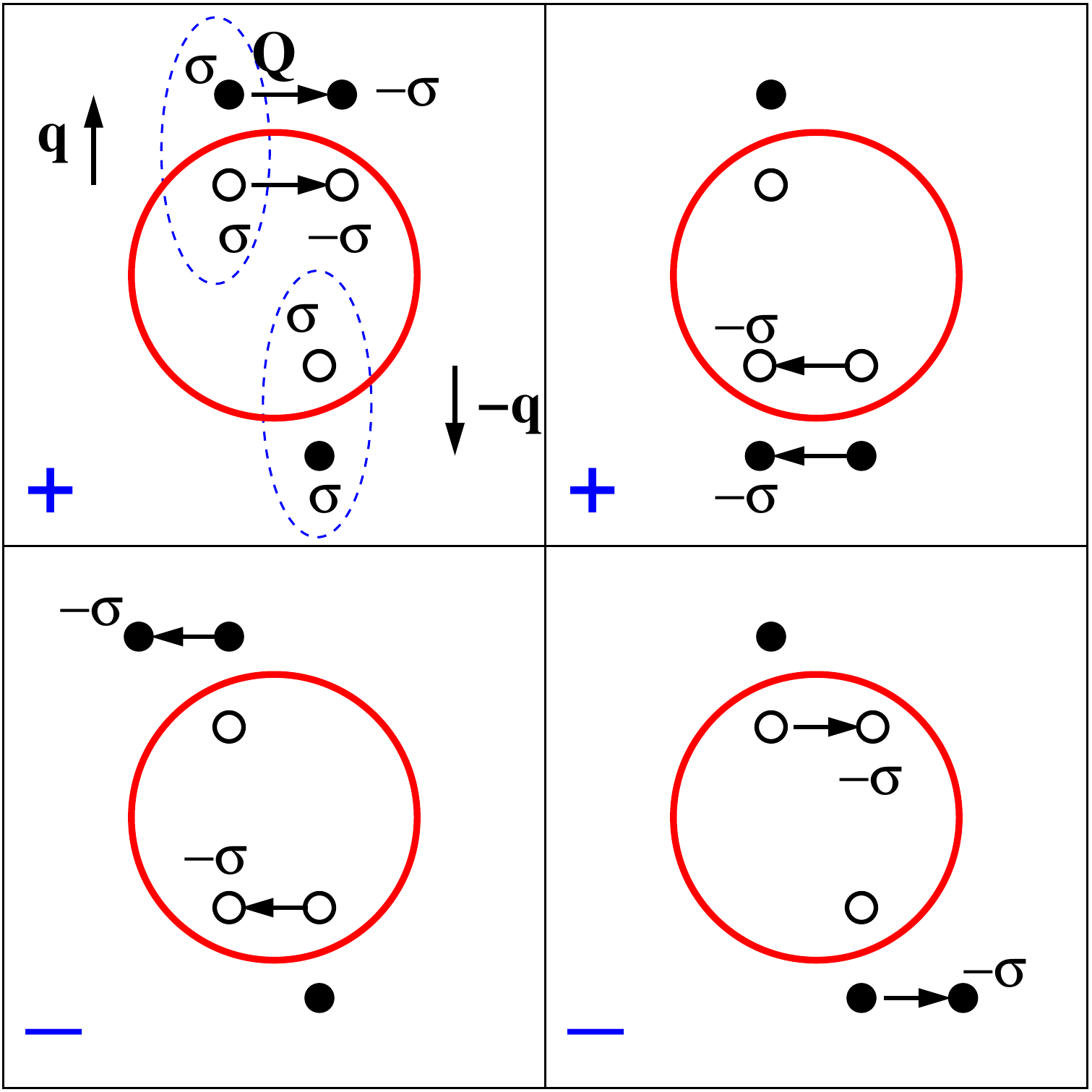}}
\caption{(Color online) Same as Fig. \ref{fig-1} but for initial pairs that have the same spin.}
\label{fig-2}
\end{figure}
If $\sigma=\sigma'$, the interaction must also flip two spins so to lead to two opposite-spin p-h pairs. The processes are shown in Fig. \ref{fig-2} with their signs. 

\begin{figure}[t]
\centerline{\includegraphics[width=7cm]{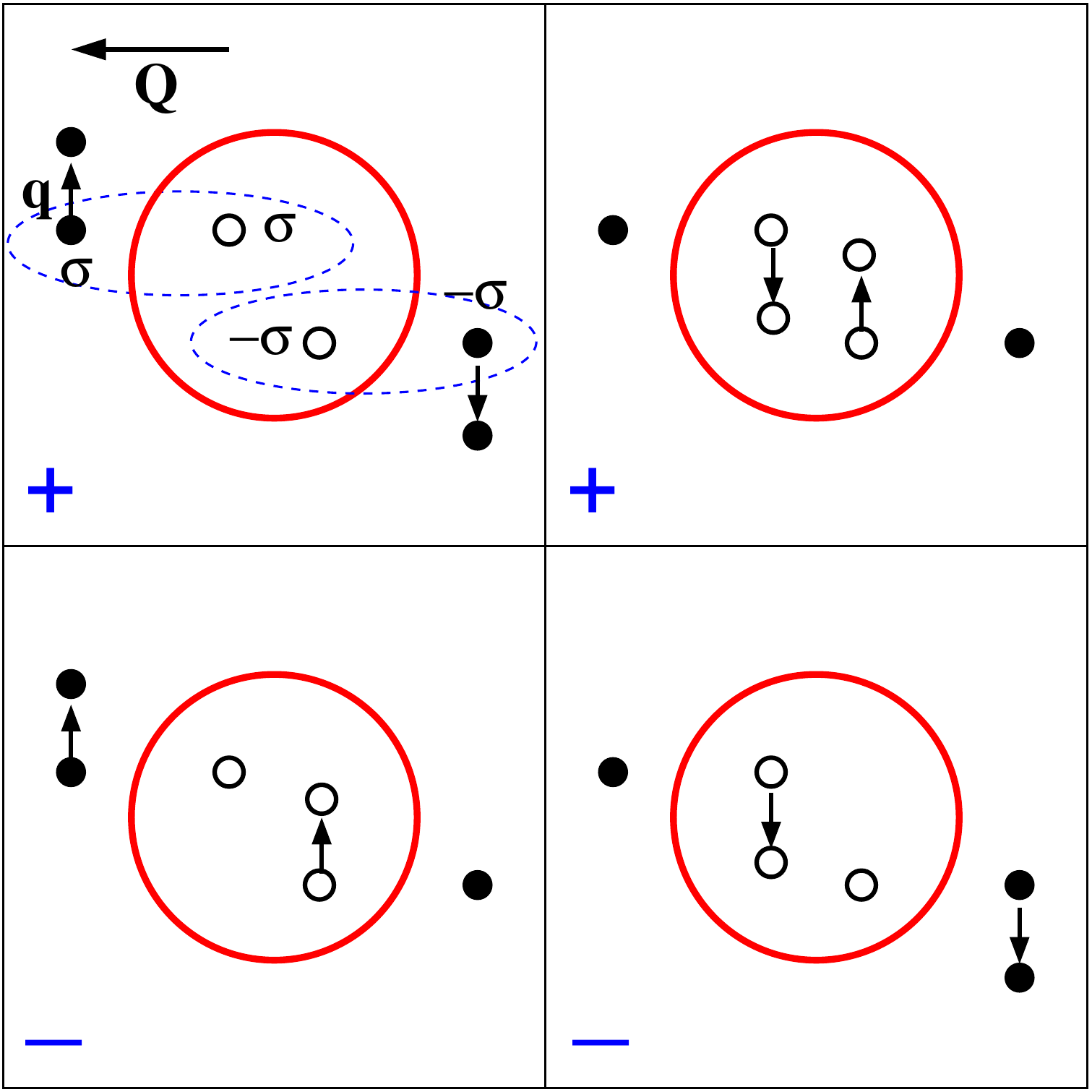}}
\caption{(Color online) Scattering processes that refer to Eq. \eqn{2nd-order-Sq-bis} and are produced by $n_{\bq\sigma}\,n_{-\bq-\sigma}$ . The sign of the process is indicated and the initial particle-hole pairs in $\ket{n}$ are encircled.}
\label{fig-3}
\end{figure}

The term in Eq. \eqn{2nd-order-Sq-bis} has contributions whenever 
two states $\ket{n}$ and $\ket{m}$, each that contains two  
p-h pairs with opposite spin and momentum, can be connected by $\mathcal{S}_\bq$. 
In particular, the processes generated by 
$n_{\bq\sigma}\,n_{-\bq-\sigma}$ are shown in Fig. \ref{fig-3},
where the states $\ket{n}$ and $\ket{m}$ have 
two p-h pairs at opposite 
momenta $\pm \bQ$ and $\pm (\bQ+\bq)$, respectively.  

\begin{figure}
\centerline{\includegraphics[width=7cm]{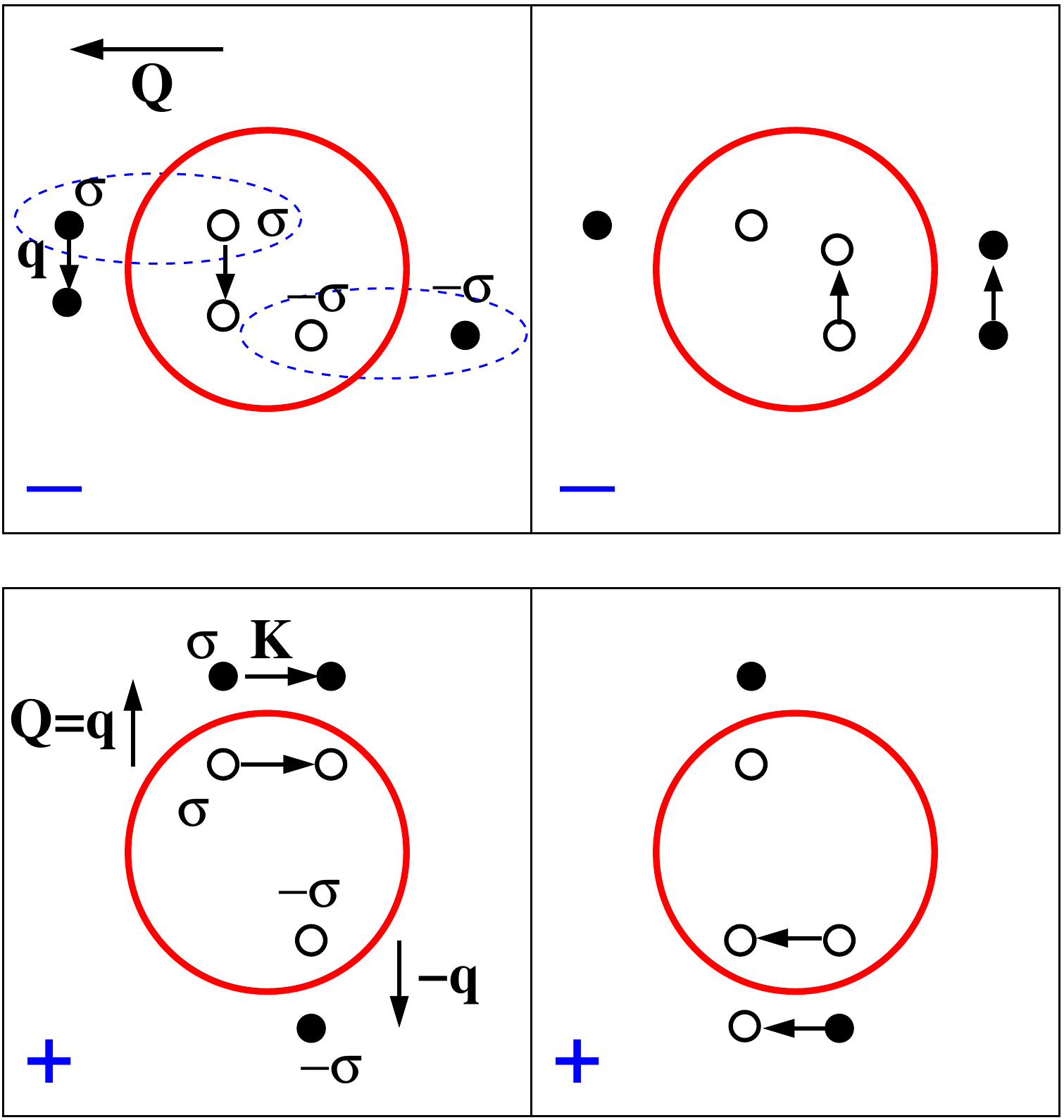}}
\caption{(Color online) Scattering processes that refer to Eq. \eqn{2nd-order-Sq-bis} and are produced by $n_{\bq\sigma}\,n_{-\bq\sigma}$ . The sign of the process is indicated and the initial particle-hole pairs in $\ket{n}$ are encircled.}
\label{fig-4}
\end{figure}

\begin{figure}[t]
\centerline{\includegraphics[width=7cm]{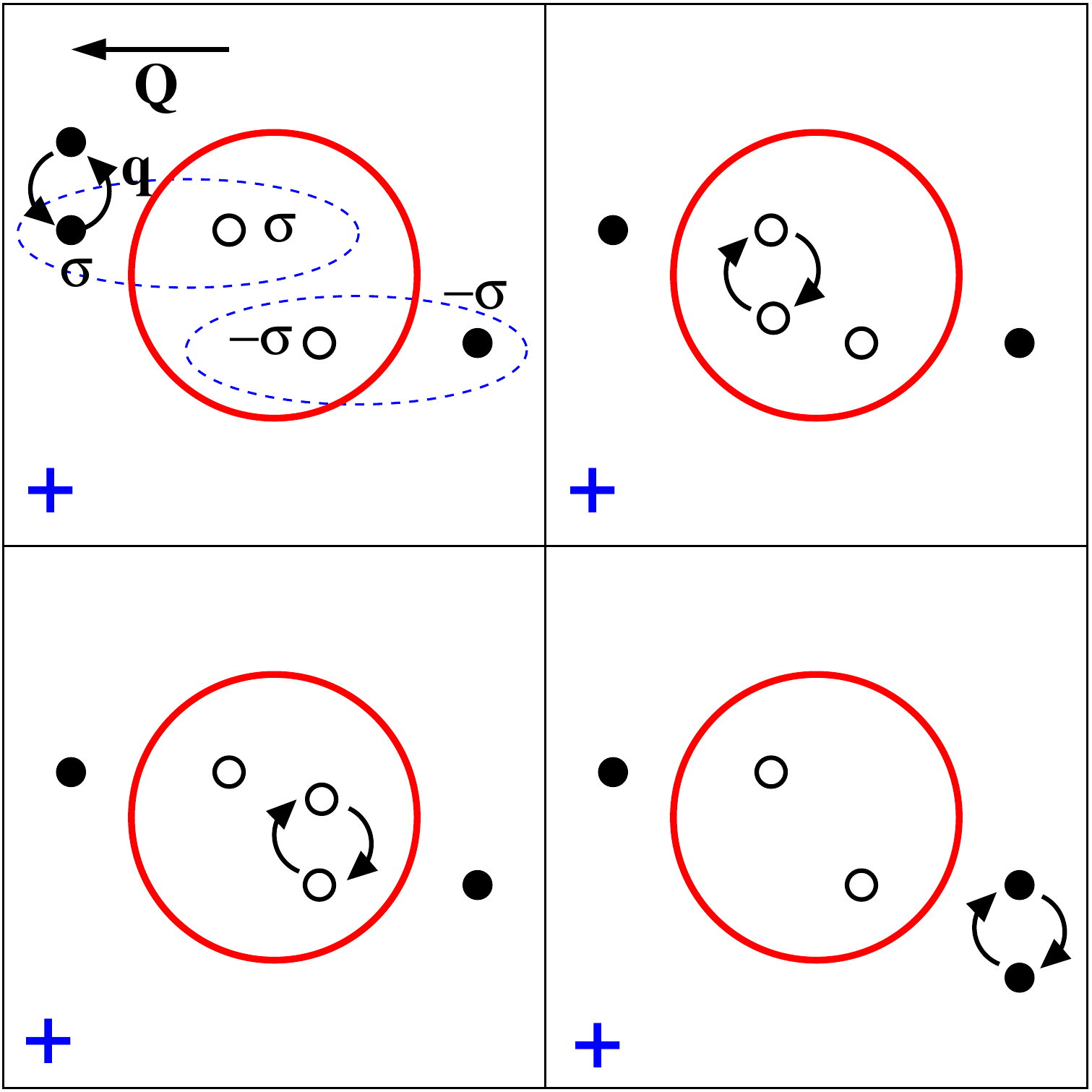}}
\caption{(Color online) Scattering processes that refer to Eq. \eqn{2nd-order-Sq-bis} and are produced by $n_{\bq\sigma}\,n_{-\bq\sigma}$ in the case in which $\ket{n}=\ket{m}$. The sign of the process is indicated and the initial particle-hole pairs in $\ket{n}$ are encircled.}
\label{fig-5}
\end{figure}

All processes that are produced by $n_{\bq\sigma}\,n_{-\bq\sigma}$ 
are instead drawn in Figs. \ref{fig-4} and \ref{fig-5}. 
In particular, the lower two panels of Fig. \ref{fig-4} refer to the case in which $\ket{n}$ has two p-h pairs already at 
momenta $\pm \bq$; the interaction simply shifts rigidly one of the pair in momentum space, in the figure the momentum shift is $\mathbf{K}$. 
 
Finally, Fig. \ref{fig-5} refers to the case in which 
$\ket{m}=\ket{n}$. Here for instance a particle is shifted by $\bq$ 
and then comes back to the initial position.

Since all states that contribute to Eqs. \eqn{2nd-order-Sq-uno} 
and \eqn{2nd-order-Sq-bis} have two p-h pairs, their linearly vanishing densities of states at low energy compensate the singularity of the denominators. Therefore the sums, which become 
integral over a continuous spectrum of degrees of freedom, are 
all convergent. Even more, all terms but one vanish as a power in $1/t$ for large times. In fact, even though 
in the sense of distributions, 
\ba
\fract{\cos Et -\cos E' t}{E' (E-E')} &\xrightarrow[t\to \infty]{~~}& -\delta(E)\,\delta(E'-E) 
+ O\left(\frac{1}{t}\right),\\
\fract{\cos (E-E')t}{E E'} &\xrightarrow[t\to \infty]{~~}& 
\delta(E)\,\delta(E')+ O\left(\frac{1}{t}\right),
\ea
those singularities are killed by the vanishing density of states of p-h excitations, leading to a null 
result for $t\to\infty$. 

The only term that actually survives is the one shown in Fig. \ref{fig-5}, which corresponds to the case 
$\ket{n}=\ket{m}$ in Eq. \eqn{2nd-order-Sq-bis}. We observe that, if we send $|\bq|\to 0$ before taking 
the limit $t\to\infty$, this term would be canceled by those shown in the upper panels of Fig. \ref{fig-4}. This simply reflects the trivial fact the $S(\bq=\bnot,t)=0$. 
The correct procedure is instead to first reach the steady-state $t\to\infty$, and only after 
send $|\bq|\to 0$. In this way we do find a finite steady-state contribution from Eqs.  
\eqn{2nd-order-Sq-uno} and \eqn{2nd-order-Sq-bis}, which reads
\bea
\delta S^{(2)}_*(\bq) &=& \frac{4U^2}{V^3}\, 
\sum_{\bk\bp\bQ}\, n_\bk\,\bar{n}_{\bk+\bQ}\,
n_{\bp+\bQ}\,\bar{n}_{\bp}\;
\Bigg\{ \label{delta-S-2nd}\\
&& 
\fract{\bar{n}_{\bk+\bQ+\bq}}{\big(\omega_{\bk+\bQ,\bk}+\omega_{\bp,\bp+\bQ}\big)
\big(\omega_{\bk+\bQ+\bq,\bk}+\omega_{\bp,\bp+\bQ}\big)}\nonumber\\
&& \!\!\!\!\!\!\!+\, \fract{n_{\bk-\bq}}{\big(\omega_{\bk+\bQ,\bk}+\omega_{\bp,\bp+\bQ}\big)
\big(\omega_{\bk+\bQ,\bk-\bq}+\omega_{\bp,\bp+\bQ}\big)}\Bigg\},\nonumber
\eea
where $n_\bk=\theta\big(k_F-|\bk|\big)$ is the momentum distribution of the Fermi sea and 
$\bar{n}_\bk=1-n_\bk$. If we now take the limit $\bq\to\bnot$, we find a term $\delta S^{(2)}_*(\bnot)$ that is finite plus a correction that starts linear in $|\bq|$. Specifically, since 
$n_\bk^2=n_\bk$, 
\bea
\delta S^{(2)}_*(\bnot) &=& \frac{8U^2}{V^3}\, 
\sum_{\bk\bp\bQ}\, \fract{n_\bk\,\bar{n}_{\bk+\bQ}\,
n_{\bp+\bQ}\,\bar{n}_{\bp}}
{\big(\omega_{\bk+\bQ,\bk}+\omega_{\bp,\bp+\bQ}\big)^2
} \label{delta-S-2nd-*-APP} \\
&=& 8\,U^2\,\int \fract{d\bQ}{(2\pi)^3}\;
\iint_0 d\omega_1 d\omega_2\, \fract{\rho_\bQ(\omega_1)\rho_\bQ(\omega_2)}{\big(\omega_1+\omega_2\big)^2},\nonumber
\eea
and is finite, i.e. not singular. The leading correction to 
$\delta S^{(2)}_*(\bnot)$ reads
\bw
\bea
\delta S^{(2)}_*(\bq)-\delta S^{(2)}_*(\bnot) &=& 
-\frac{8U^2}{V^3}\, 
\sum_{\bk\bp\bQ}\, \fract{\bar{n}_{\bk+\bQ}\,
n_{\bp+\bQ}\,\bar{n}_{\bp}}
{\big(\omega_{\bk+\bQ,\bk}+\omega_{\bp,\bp+\bQ}\big)
\big(\omega_{\bk+\bQ,\bk-\bq}+\omega_{\bp,\bp+\bQ}\big)}\; n_\bk\,\bar{n}_{\bk-\bq}\nonumber\\
&\simeq& -|\bq|\,\fract{U^2}{4}\, \int \fract{d\bQ}{(2\pi)^3}
\,\fract{1}{Q}\; \int_{\text{Max}\big[0,Q(Q-2)\big]}^{Q(2+Q)}
d\omega_1\,\int_0 d\omega_2\,\rho_\bQ(\omega_2)\,
\fract{1}{\big(\omega_1+\omega_2\big)^2}+ O(q^2).
\eea
\ew
Therefore the second order correction to the coefficient of the non-analytic term $\propto |\bq|$ in the steady-state differs from that at equilibrium. 

\section{Diagonalization of the effective Hamiltonian for the p-p and h-h excitations in the low density limit}
\label{App:sec3}

In this section we explicitly diagonalize the Hamiltonian 
\bea
\mathcal{H}_* &=& 
\sum_\bQ\,\mathcal{H}_\bQ = \sum_\bQ\,
\bigg[
\sum_{\bk}\,\omega_{\bk,\bQ}\,\Big(a^\dagger_{\bk,\bQ}a^\dagga_{\bk,\bQ}  + 
b^\dagger_{\bk,\bQ}b^\dagga_{\bk,\bQ}\Big) \nonumber \\
&& \!\!\! + 
\fract{U}{V}\,\sum_{\bk,\bp}\,\Big(b^\dagger_{\bk\bQ} - a^\dagga_{\bk,\bQ}\Big)
\Big(b^\dagga_{\bp\bQ} - a^\dagger_{\bp,\bQ}\Big)
\bigg],   \label{SM-H-PWA}
\eea
relaxing the hard-core constraint. In Eq. \eqn{SM-H-PWA} 
$b^\dagger_{\bk,\bQ}$ creates two particles outside the Fermi sea with momenta $\bk$, spin $\up$ and $-\bk+\bQ$, spin $\down$, i.e. 
$\bQ$ is the total momentum of the pair. On the contrary, 
$a^\dagger_{\bk,\bQ}$ creates two holes within the Fermi sea, 
one at momentum $\bk$, spin $\up$, and the other at momentum $-\bk+\bQ$, spin $\down$. We recall that $\mathcal{H}_*$ describes the excitations of the Fermi sea brought by the interaction in the dilute limit. Hereafter we shall use dimensionless units in which 
momentum is in units of $k_F$, energy in units of 
$\epsilon_F = \hbar^2 k_F^2/2m$, and time in units of $\hbar/\epsilon_F$. In addition, the constraint of being outside or inside the Fermi sea will be implicitly hidden in the definition of $b^\dagga_{\bk,\bQ}$ and $a^\dagga_{\bk,\bQ}$, respectively.   
 
We start by noting that a pair of holes requires $Q\leq 2$ ($2k_F$ in dimensional units), so that 
\bea
H_* &=& \sum_{\bk,\bQ:|\bQ|\leq 2} \! \omega_{\bk,\bQ}\,a^\dagger_{\bk,\bQ}a^\dagga_{\bk,\bQ}
+ \omega_{\bk,\bQ}\,b^\dagger_{\bk,\bQ}b^\dagga_{\bk,\bQ} 
\nonumber\\
&& + \frac{U}{V}\,\sum_{\bk,\bp,\bQ:|\bQ|\leq 2}  \!
\Big(b^\dagger_{\bp,\bQ}-a^\dagga_{\bp,\bQ}\Big)\Big(b^\dagga_{\bk,\bQ}-a^\dagger_{\bk,\bQ}\Big)\label{H*} \\
&& + \sum_{\bk,\bQ:|\bQ|> 2} \! \omega_{\bk,\bQ}\,b^\dagger_{\bk,\bQ}b^\dagga_{\bk,\bQ}
 + \frac{U}{V} \sum_{\bk,\bp,\bQ:|\bQ|> 2}  \!
b^\dagger_{\bp,\bQ}b^\dagga_{\bk,\bQ}.\nonumber
\eea
The last two terms are diagonalized by a simple unitary transformation, while the former two by a generalized canonical transformation, which we shall focus on first. 

Let us therefore 
diagonalize $H_*$ in a subspace at fixed $\bQ$ such that $Q\leq 2$. For simplicity we shall drop the 
label $\bQ$. We define the canonical transformation
\ba
a^\dagga_{\bk} &=& \sum_\ep\,U_{\bk\ep}\,\alpha^\dagga_\ep + \sum_\bep\,V_{\bk\bep}\,\beta^\dagger_\bep,\\
b^\dagger_{\bk} &=& \sum_\bep\,Z_{\bk\bep}\,\beta^\dagger_\bep + \sum_\ep\,W_{\bk\ep}\,\alpha^\dagga_\ep,
\ea
where $\hat{U}$ and $\hat{Z}$ are square real matrices, while $\hat{V}$ and $\hat{W}$ are still real but in general rectangular 
-- the number of holes being much smaller than the number of particles in the low-density limit -- 
satisfying
\ba
U\,U^T - V\,V^T &=& I,\\
Z\,Z^T - W\,W^T &=& I,\\
U\,W^T - V\,Z^T &=& 0,\\
U^T\,U - W^T\,W &=& I,\\
Z^T\,Z - V^T\,V &=& I,\\
U^T\,V - W^T\,Z &=& 0.
\ea
The inverse transformation then reads
\ba
\alpha^\dagga_{\ep} &=& \sum_\bk\,U_{\bk\ep}\,a^\dagga_\bk - \sum_\bk\,W_{\bk\ep}\,
b^\dagger_\bk,\\
\beta^\dagger_{\bep} &=& \sum_\bk\,Z_{\bk\bep}\,b^\dagger_\bk - \sum_\bk\,V_{\bk\bep}\,a^\dagga_\bk.
\ea
We define the matrix $A$ with elements
\[
A_{\bk\bp} = \omega_{\bk}\,\delta_{\bk\bp} + \frac{U}{V},
\]
for $|\bk|\leq 1$ and $|-\bk+\bQ|\leq 1$ ($\leq k_F$ in dimensional units),  
the matrix $B$ with elements
\[
B_{\bk\bp} = \omega_{\bk}\,\delta_{\bk\bp} + \frac{U}{V},
\]
for $|\bk| > 1$ and $|-\bk+\bQ|> 1$, 
and finally the matrix $D$ with elements
\[
D_{\bk\bp} = - \frac{U}{V},
\]
which couples the interior to the exterior of the Fermi sphere. The interaction $U$ is also measured in units of $\ep_F$.  
With the above definitions, the diagonalization of the Hamiltonian corresponds to the eigenvalue equation
\[
\begin{pmatrix}
A & D\\
-D^T & -B
\end{pmatrix}\,
\begin{pmatrix}
U & V\\
W & Z
\end{pmatrix} = \begin{pmatrix}
U & V\\
W & Z
\end{pmatrix}
\begin{pmatrix}
\epsilon & 0\\
0 & -\bep
\end{pmatrix}
\]
The solution can be readily found. Specifically, the eigenvalues satisfy
\bea
\! 1 \! &=& \! \frac{U}{V}\,\sum^{\text{IN}}_{\bk}\, \fract{1}{\epsilon-\omega_\bk} 
- \frac{U}{V}\,\sum^{\text{OUT}}_{\bk}\, \fract{1}{\epsilon+\omega_\bk}
\equiv U\,\chi(-\epsilon),\label{eigen-1}\\
\! 1 \! &=& \! \frac{U}{V}\,\Sout_\bk\, \fract{1}{\bep-\omega_\bk} 
- \frac{U}{V}\,\Sin_\bk\, \fract{1}{\bep+\omega_\bk} \equiv
U\,\chi(\bep),\;\;\label{eigen-2}
\eea 
where the suffix IN means $|\bk|\leq 1$ and $|-\bk+\bQ|\leq 1$, while 
OUT refers to $|\bk|>1$ and $|-\bk+\bQ|>1$. 
In fact, one can solve for any $x\lessgtr 0$,
\be
U\,\chi(x) = \frac{U}{V}\,\Sout_\bk\, \fract{1}{x-\omega_\bk} 
-  \frac{U}{V}\,\Sin_\bk\,\fract{1}{x+\omega_\bk} = 1,\label{eigen}
\ee
and set the positive solutions to $\bep$ and the negative ones to $-\epsilon$.
The coefficients of the canonical transformation read
\ba
U_{\bk\epsilon} &=& \sqrt{\fract{U}{V}}\; N_\epsilon\; \fract{1}{\epsilon-\omega_\bk},\\
W_{\bk\epsilon} &=& \sqrt{\fract{U}{V}}\;N_\epsilon\; \fract{1}{\epsilon+\omega_\bk},\\
Z_{\bk\bep} &=& \sqrt{\fract{U}{V}}\;N_\bep\; \fract{1}{\bep-\omega_\bk},\\
V_{\bk\bep} &=& \sqrt{\fract{U}{V}}\;N_\bep\; \fract{1}{\bep+\omega_\bk},\\
\ea
with the parameters $N_\epsilon$ and $N_\bep$ that should be determined by imposing the transformation 
to be indeed canonical, i.e.
\ba
1 &=& N_\epsilon^2\;\frac{U}{V}\,\Sin_\bk\, \fract{1}{\big(\epsilon-\omega_\bk\big)^2}
- N_\epsilon^2\;\frac{U}{V}\,\Sout_\bk\, \fract{1}{\big(\epsilon+\omega_\bk\big)^2},\\
1 &=& N_\bep^2\;\frac{U}{V}\,\Sout_\bk\, \fract{1}{\big(\bep-\omega_\bk\big)^2}
- N_\bep^2\;\frac{U}{V}\,\Sin_\bk\, \fract{1}{\big(\bep+\omega_\bk\big)^2},\\
0 &=& N_\epsilon\,N_\bep\,\Sin_\bk\, \fract{1}{\big(\epsilon-\omega_\bk\big)\big(
\bep+\omega_\bk\big)} \\
&& - N_\epsilon\,N_\bep\,\Sout_\bk\, \fract{1}{\big(\bep-\omega_\bk\big)\big(\epsilon+\omega_\bk\big)}.
\ea
We observe that the last condition is also equivalent to 
\ba
0 &=& N_\epsilon\,N_\bep\,\Sin_\bk\, \fract{1}{\big(\epsilon-\omega_\bk\big)\big(
\bep+\omega_\bk\big)} \\
&& - N_\epsilon\,N_\bep\,\Sout_\bk\, \fract{1}{\big(\bep-\omega_\bk\big)\big(\epsilon+\omega_\bk\big)}\\
&=& V\,\fract{N_\epsilon\,N_\bep}{\epsilon+\bep}\,\Big[ \chi(-\epsilon) - \chi(\bep)\Big] = 0,
\ea
hence is automatically satisfied because of the eigenvalue equations \eqn{eigen-1} and \eqn{eigen-2}.

We conclude by noting that the same calculation can be carried out also for $Q>2$. In this case there are no $a$-bosons, and one only needs to find the unitary transformation $Z_{\bk\bep}$, i.e. one has to solve
\be
1 = \frac{U}{V}\,\Sout_\bk\, \fract{1}{\bep-\omega_\bk},\label{eigen-2-Q>2}
\ee
and define 
\[
Z_{\bk\bep} = \sqrt{\fract{U}{V}}\; N_\bep \; \fract{1}{\bep-\omega_\bk},
\]
where
\[
N_\bep^{-2} = \frac{U}{V}\,\Sout_\bk\,\fract{1}{\big(\bep-\omega_\bk\big)^2}.
\]

\subsection{Continuum limit}
\label{Continuum limit}
A proper definition of a steady-state requires to take first the 
thermodynamic limit, hence to turn all finite sums into integrals over a continuum of degrees of freedom, and only afterwards send the time 
$t\to\infty$. 

In order to understand what the above formulas mean in the continuum limit, we may follow the different route to solve the problem at equilibrium within the Matsubara technique. We would find for instance that 
the imaginary-time Fourier transforms
 of $G_b(\tau,\bk) = -\langle T_\tau\big(b^\dagga_\bk(\tau)\,b^\dagger_\bk\big)\rangle$
and $G_a(\tau,\bk) = -\langle T_\tau\big(a^\dagga_\bk(\tau)\,a^\dagger_\bk\big)\rangle$ read
\ba
G_b(i\Omega,\bk) &=& \fract{1}{i\Omega-\omega_\bk} + 
\frac{1}{V}\, \fract{T(i\Omega)}{\big(i\Omega-\omega_\bk\big)^2},\\
G_a(i\Omega,\bk) &=& \fract{1}{i\Omega-\omega_\bk} + 
\frac{1}{V}\, \fract{T(-i\Omega)}{\big(i\Omega-\omega_\bk\big)^2},
\ea
where 
\be
T(i\Omega) = \fract{U}{1- U\,\chi(i\Omega)},\label{matrice-T}
\ee
is the usual definition of the $T$-matrix, with $\chi(z)$ the Cooper bubble that actually corresponds 
to the same function defined above continued in the complex frequency plane. In the continuum limit $\chi(z)$ has a branch cut on the real axis, specifically 
\bw
\ba
\chi(\ep + i0^+) - \chi(\ep - i0^+) &=& 2\pi i\,\frac{1}{V}\,\Sin_\bk\,\delta(\ep+\omega_\bk)
- 2\pi i\, \frac{1}{V}\,\Sout_\bk\,\delta(\ep-\omega_\bk)\equiv 2\pi i\, \mathcal{N}_\text{IN}(-\ep)
- 2\pi i\,\mathcal{N}_\text{OUT}(\epsilon),
\ea
\ew
where $\mathcal{N}_\text{IN}(x)$ and $\mathcal{N}_\text{OUT}(x)$ are defined only for 
$x>0$ and correspond to the density of states of pairs of holes and particles, respectively, at total 
momentum $\bQ$. Introducing back the dependence on $\bQ$ and in dimensionless units, 
\bw
\bea
\fract{1}{\left(k_F a\right)^3}\; \mathcal{N}_\text{IN}(\ep,\bQ) &=& \theta\Big(Q(2-Q)-\ep\Big)\; \fract{\ep}{16\pi^2\,Q}
\nonumber\\
&& + \theta\Big(\ep-Q(2-Q)\Big)\,\theta\Big((2+Q)(2-Q)-2\ep\Big)\; 
\fract{\sqrt{4-2\ep-Q^2}}{16\pi^2 },\label{N-IN}\\
\fract{1}{\left(k_F a\right)^3}\; \mathcal{N}_\text{OUT}(\ep,\bQ\leq 2) &=& \theta\Big(Q(2+Q)-\ep\Big)\; \fract{\ep}{16\pi^2\,Q}
+ \theta\Big(\ep-Q(2+Q)\Big)\; 
\fract{\sqrt{4+2\ep-Q^2}}{16\pi^2},\label{N-OUT-Q<2}\\
\fract{1}{\left(k_F a\right)^3}\; \mathcal{N}_\text{OUT}(\ep,\bQ>2) &=& 
\theta\Big(2\ep-(Q+2)(Q-2)\Big)\, \theta\Big(Q(Q-2)-\ep\Big)\;
\fract{\sqrt{2\ep+4-Q^2}}{16\pi^2} \nonumber\\
&& + \theta\Big(\ep-Q(2-Q)\Big)\theta\Big(Q(2+Q)-\ep\Big)\;\fract{\ep}{16\pi^2 Q}\nonumber \\
&& + \theta\Big(\ep-Q(2+Q)\Big)\; \fract{\sqrt{2\ep+4-Q^2}}{16\pi^2},
\label{N-OUT-Q>2}
\eea
where $a$ is the lattice spacing.

We then note that, in the limit of zero temperature,  
\ba
\langle b^\dagger_\bk b^\dagga_\bk\rangle &=& -T\sum_{\Omega}\,\text{e}^{i\Omega 0^+}\,
G_b(i\Omega,\bk) 
= \frac{U}{V}\,\int_{0}^\infty\,d\epsilon\;\fract{1}{\big(\ep+\omega_\bk\big)^2}\;
\fract{U\,\mathcal{N}_\text{IN}(\ep)}{\big(1-U\,\chi'(-\ep)\big)^2 + \pi^2\, U^2\,\mathcal{N}_\text{IN}(\ep)^2}
,\\
\langle a^\dagger_\bk a^\dagga_\bk\rangle &=& -T\sum_{\Omega}\,\text{e}^{i\Omega 0^+}\,
G_a(i\Omega,\bk) 
= \frac{U}{V}\,\int_{0}^\infty\,d\epsilon\;\fract{1}{\big(\ep+\omega_\bk\big)^2}\;
\fract{U\,\mathcal{N}_\text{OUT}(\ep)}{\big(1-U\,\chi'(\ep)\big)^2 + \pi^2\, U^2\,\mathcal{N}_\text{OUT}(\ep)^2},
\ea
\ew
where $\chi'(\ep) = \Re\text{e}\,\chi(\ep-i0^+)$.
On the other hand, if we calculate the above average values directly via exact diagonalization, we find 
\ba
\langle b^\dagger_\bk b^\dagga_\bk\rangle &=& \frac{U}{V}\, \sum_{\ep}\, 
N_\ep^2\,\fract{1}{\big(\ep+\omega_\bk\big)^2},\\
\langle a^\dagger_\bk a^\dagga_\bk\rangle &=& \frac{U}{V}\, \sum_{\bep}\, 
N_\bep^2\,\fract{1}{\big(\bep+\omega_\bk\big)^2},
\ea
showing that, in the continuum limit $\sum_\ep \to \int d\ep$, 
\ba
U\,N_\ep^2 &\to&\frac{1}{\pi}\, \Im\text{m}\, T(-\epsilon+i0^+) = N_\text{IN}(\ep),\\
U\,N_\bep^2 &\to&\frac{1}{\pi}\, \Im\text{m}\, T(\bep-i0^+)= N_\text{OUT}(\bep). 
\ea

\subsection{Time dependent averages}
\label{Time dependent averages}

The advantage of the exact diagonalization is to allow calculating the out-of-equilibrium evolution after 
suddenly switching on $U$ at time $t=0$, without solving any integral equation. The initial state is thence the vacuum of the original bosons, but the operators are time-evolved with the $U\not=0$ Hamiltonian.  
By means of the exact-diagonalization, we find that, 
for $Q\leq 2$, 
\ba
\langle \alpha^\dagger_{\ep}(t)\alpha^\dagga_{\ep'}(t)\rangle &=& N_{\ep}\,N_{\ep'}\, 
\text{e}^{i(\ep-\ep')t}\; 
\frac{U}{V}
\Sout_\bp \fract{1}{\big(\ep+\omega_\bp\big)\big(\ep'+\omega_\bp\big)},\\
\langle \beta^\dagger_{\bep}(t)\beta^\dagga_{\bep'}(t)\rangle &=& N_{\bep}\,N_{\bep'}\, 
\text{e}^{i(\bep-\bep')t}\; 
\frac{U}{V}
\Sin_\bp \fract{1}{\big(\bep+\omega_\bp\big)\big(\bep'+\omega_\bp\big)},\\
\langle \alpha^\dagger_{\ep}(t)\beta^\dagger_{\bep}(t)\rangle &=& -N_{\ep}\,N_{\bep}\, 
\text{e}^{i(\ep+\bep)t}\; 
\frac{U}{V}
\Sout_\bp \fract{1}{\big(\ep+\omega_\bp\big)\big(\bep-\omega_\bp\big)},\\
\langle \beta^\dagger_{\bep}(t)\alpha^\dagger_{\ep}(t)\rangle &=& -N_{\ep}\,N_{\bep}\, 
\text{e}^{i(\ep+\bep)t}\; 
\frac{U}{V}
\Sin_\bp \fract{1}{\big(\ep-\omega_\bp\big)\big(\bep+\omega_\bp\big)}. 
\ea
All other averages can be obtained by the above ones, as for instance
\[
\langle \alpha^\dagga_{\ep'}(t) \alpha^\dagger_{\ep}(t)\rangle 
= \delta_{\ep\ep'} + \langle \alpha^\dagger_{\ep}(t)\alpha^\dagga_{\ep'}(t)\rangle,
\]
or
\[
\langle \alpha^\dagga_{\ep}(t)\beta^\dagga_{\bep}(t)\rangle = \Big(\langle \beta^\dagger_{\bep}(t)\alpha^\dagger_{\ep}(t)\rangle\Big)^*.
\]
For $Q>2$ the boson vacuum remains instead unaffected, only the excitation energies are modified by interaction.

It follows therefore that, for $Q\leq 2$, 
\bw
\bea
\langle a^\dagger_\bk(t) a^\dagga_\bk(t)\rangle &=&  \frac{U}{V}\,\sum_{\bep}\, N_{\bep}^2\,\fract{1}{\big(\bep+\omega_\bk\big)^2} \nonumber \\
&& + \fract{U^2}{V^2}\, \sum_{\ep\ep'}\, 
\cos\,(\ep-\ep')t\; N_{\ep}^2\,N_{\ep'}^2\, 
\fract{1}{\big(\ep-\omega_\bk\big)\big(\ep'-\omega_\bk\big)}\, \Sout_\bp\, \fract{1}{\big(\ep+\omega_\bp\big)\big(\ep'+\omega_\bp\big)}\nonumber\\
&& 
+ \fract{U^2}{V^2}\,\sum_{\bep\bep'}\,  
\cos(\bep-\bep')t\; N_{\bep}^2\,N_{\bep'}^2\, 
\fract{1}{\big(\bep+\omega_\bk\big)\big(\bep'+\omega_\bk\big)} \; 
\Sin_\bp\, \fract{1}{\big(\bep+\omega_\bp\big)\big(\bep'+\omega_\bp\big)}\nonumber\\
&& - \frac{U^2}{V^2}\,\sum_{\ep\bep}\, 2\cos\big(\ep t +\bep t\big)\, 
N_\ep^2\,N_\bep^2\, \fract{1}{\big(\ep-\omega_\bk\big)\big(\bep+\omega_\bk\big)}\; 
\Sout_\bp\, \fract{1}{\big(\ep+\omega_\bp\big)\big(\bep-\omega_\bp\big)},\label{aa}\\
\langle b^\dagger_\bk(t) b^\dagga_\bk(t)\rangle &=& 
 \frac{U}{V}\,\sum_{\ep}\, N_{\ep}^2\,\fract{1}{\big(\ep+\omega_\bk\big)^2}\nonumber\\
&& + \fract{U^2}{V^2}\, \sum_{\bep\bep'}\, 
\cos(\bep-\bep')t\; N_{\bep}^2\,N_{\bep'}^2\, 
\fract{1}{\big(\bep-\omega_\bk\big)\big(\bep'-\omega_\bk\big)}\, \Sin_\bp\, \fract{1}{\big(\bep+\omega_\bp\big)\big(\bep'+\omega_\bp\big)}\nonumber\\
&&  
+ \fract{U^2}{V^2}\,\sum_{\ep\ep'}\,  
\cos(\ep-\ep')t\; N_{\ep}^2\,N_{\ep'}^2\, 
\fract{1}{\big(\ep+\omega_\bk\big)\big(\ep'+\omega_\bk\big)} \; 
\Sout_\bp\, \fract{1}{\big(\ep+\omega_\bp\big)\big(\ep'+\omega_\bp\big)}\nonumber\\
&& - \frac{U^2}{V^2}\,\sum_{\ep\bep}\, 2\cos\big(\ep t +\bep t\big)\, 
N_\ep^2\,N_\bep^2\, \fract{1}{\big(\bep-\omega_\bk\big)\big(\ep+\omega_\bk\big)}\; 
\Sin_\bp\, \fract{1}{\big(\bep+\omega_\bp\big)\big(\ep-\omega_\bp\big)},\label{bb}
\eea
\ew
while, for $Q>2$, $\langle b^\dagger_\bk(t) b^\dagga_\bk(t)\rangle = 0$. 

The first terms on the right hand sides of Eqs. \eqn{aa} and \eqn{bb} are the equilibrium values, hence all the rest is due to the sudden quench. We observe that, because of the eigenvalue equations, 
\bw
\bea
\frac{U}{V}\,\Sin_\bp\, \fract{1}{\big(\bep+\omega_\bp\big)\big(\ep-\omega_\bp\big)} &=& 
\frac{U}{V}\Sout_\bp\, \fract{1}{\big(\ep+\omega_\bp\big)\big(\bep-\omega_\bp\big)}
= \frac{1}{\ep+\bep}\,\bigg[
1 + \frac{U}{V}\,\Sin_\bp\, \fract{1}{\bep+\omega_\bp}
+ \frac{U}{V}\Sout_\bp\, \fract{1}{\ep+\omega_\bp}\bigg],\label{D}
\eea
\ew
which therefore brings no singularity when $\bep=\omega_\bp$. 

Since before the continuum limit is taken $T(\omega_\bk)=0$, if we consider a contour that run anti-clockwise closely around the positive real axis, then
\bw
\ba
\bar{I}=U^2\,\sum_{\bep\bep'}\, N_\bep^2\,N_{\bep'}^2\, 
\fract{\cos(\bep t -\bep' t)}{\big(\bep-\omega_\bk\big)\big(\bep'-\omega_\bk\big)}\;
F(\bep,\bep') 
&=& \oint \fract{dz\,dz'}{(2\pi i)^2}\;T(z)\,T(z')\,
\fract{\cos(z t -z' t)}{\big(z-\omega_\bk\big)\big(z'-\omega_\bk\big)}\;
F(z,z'),
\ea
where $F(z,z')=F(z',z)$ is assumed analytic. We can now take the continuum limit and find that 
\ba
\bar{I} &\to& 
\fint_0 d\ep\,d\ep'\,N_\text{OUT}(\ep)\,N_\text{OUT}(\ep')\,
\fract{\cos(\ep t -\ep' t)}{\big(\ep-\omega_\bk\big)\big(\ep'-\omega_\bk\big)}\;
F(\ep,\ep')\\
&& - 2T'(\omega_\bk)\,\fint d\ep\,N_\text{OUT}(\ep)\,
\fract{\cos(\ep t -\omega_\bk t)}{\big(\ep-\omega_\bk\big)}\;F(\ep,\omega_\bk)
+ T'(\omega_\bk)^2\,F(\omega_\bk,\omega_\bk),
\ea 
\ew
where $\fint d\ep \,(\dots)$ means the 
Cauchy principal value integration and 
$T'(\ep)= \Re\text{e}\,T(\ep-i0^+)$. Seemingly,
\bw
\ba
I&=&U^2\,\sum_{\ep\ep'}\, N_\ep^2\,N_{\ep'}^2\, 
\fract{\cos(\ep t -\ep' t)}{\big(\ep-\omega_\bk\big)\big(\ep'-\omega_\bk\big)}\;
F(\ep,\ep')
=\oint \fract{dz\,dz'}{(2\pi i)^2}\;T(-z)\,T(-z')\,
\fract{\cos(z t -z' t)}{\big(z-\omega_\bk\big)\big(z'-\omega_\bk\big)}\;
F(z,z')\\
&\to& \fint_0 d\ep\,d\ep'\,N_\text{IN}(\ep)\,N_\text{IN}(\ep')\,
\fract{\cos(\ep t -\ep' t)}{\big(\ep-\omega_\bk\big)\big(\ep'-\omega_\bk\big)}\;
F(\ep,\ep')- 2T'(-\omega_\bk)\,\fint d\ep\,N_\text{IN}(\ep)\,
\fract{\cos(\ep t -\omega_\bk t)}{\big(\ep-\omega_\bk\big)}\;F(\ep,\omega_\bk) \\
&& \qquad\qquad\qquad \qquad\qquad\qquad+ T'(-\omega_\bk)^2\,F(\omega_\bk,\omega_\bk).
\ea 
\ew
We are actually interested in the large-$t$ limit. We observe that
\be
\lim_{t\to\infty}\, \fract{\sin(\ep t -\omega_\bk t)}{\ep-\omega_\bk}
= \pi\,\delta\big(\ep-\omega_\bk\big),
\ee
so that
\bw
\bea
\lim_{t\to\infty}\,\bar{I}  &=& \Big[T'(\omega_\bk)^2 + \pi^2\,N_\text{OUT}(\omega_\bk)^2\Big]\,F(\omega_\bk,\omega_\bk)= \Big|T(\omega_\bk-i0^+)\Big|^2\,F(\omega_\bk,\omega_\bk)
,\label{bar-I}
\eea
and 
\bea
\lim_{t\to\infty}\,I &=& \Big[T'(-\omega_\bk)^2 + \pi^2\,N_\text{IN}(\omega_\bk)^2\Big]\,F(\omega_\bk,\omega_\bk)=\Big|T(-\omega_\bk+i0^+)\Big|^2\,F(\omega_\bk,\omega_\bk).\label{I}
\eea
\ew
\subsection{Steady state values}
\label{Steady state values}

We are now in the position to evaluate the steady state value 
of the Fermi distribution jump $Z(t)$ defined by 
\be
Z(t) \simeq 1 - \sum_\bQ\, \bra{\psi(t)} a^\dagger_{\bk,\bQ}a^\dagga_{\bk,\bQ}
+ b^\dagger_{\bk,\bQ}b^\dagga_{\bk,\bQ}\ket{\psi(t)} 
,\label{SM-Z-expression}
\ee
where $|\bk|=k_F$ and $\ket{\psi(t)}$ is the boson-vacuum evolved with the Hamiltonian \eqn{SM-H-PWA}. Once we set $|\bk|=k_F$ and integrate over $\bQ$, 
the only terms in Eqs. \eqn{aa} and \eqn{bb} than could survive in the $t\to\infty$ limit are, apart from the equilibrium values, the second ones. By the formulas above, we thence find, inserting back the total momentum label $\bQ$, that  
\bw
\bea
\langle a^\dagger_{\bk,\bQ}(t) a^\dagga_{\bk,\bQ}(t)\rangle &\to& 
\frac{1}{V}\, \int_0 \fract{d\ep}{\pi}\,
\fract{\Im\text{m} T\big(\ep-i0^+,\bQ\big)}{\big(\ep+\omega_{\bk,\bQ}\big)^2}
+ \frac{1}{V}\,\Big|T\big( \!-\omega_{\bk,\bQ}+i0^+,\bQ\big)\Big|^2\, \int\,d\omega\, 
\fract{\mathcal{N}_\text{OUT}(\omega,\bQ)}{\big(\omega_{\bk,\bQ}+\omega\big)^2},\label{aa-steady}\\
\langle b^\dagger_{\bk,\bQ}(t) b^\dagga_{\bk,\bQ}(t)\rangle &\to& 
\frac{1}{V}\, \int_0 d\ep\,\fract{\Im\text{m} T\big( \! -\ep+i0^+,\bQ\big)}{\big(\ep+\omega_{\bk,\bQ}\big)^2}
 + \frac{1}{V}\,\Big|T\big(\omega_{\bk,\bQ}-i0^+,\bQ\big)\Big|^2\, 
\int\,d\omega\, 
\fract{\mathcal{N}_\text{IN}(\omega,\bQ)}{\big(\omega_{\bk,\bQ}+\omega\big)^2} \label{bb-steady}.
\eea
\ew
As anticipated, the boson occupation numbers are $\sim 1/V$, thus justifying our discarding the hard-core constraint. 

Through  
Eqs. \eqn{SM-Z-expression} -- \eqn{bb-steady} above, and 
by means of Eqs. \eqn{A2} and \eqn{A4} below, 
we find that the 
steady state value of $Z_* = Z(t\to\infty)$ reads
\bw
\bea
Z_* &=& Z_\text{eq.} - 
\fract{(k_F a)^3}{8\pi^2}\,\int_0^2 \,Q\,dQ\,
\Bigg\{ \int_0^{Q(2-Q)}\,d\omega\,
\Big|T\big(\! -\omega+i0^+,\bQ\big)\Big|^2\, \int_0\,d\ep\, 
\fract{\mathcal{N}_\text{OUT}(\ep,\bQ)}{\big(\omega +\ep\big)^2}\,\Bigg]\nonumber\\
&& + \int_0^{Q(2+Q)}\,d\omega\,
\Big|T\big(\omega-i0^+,\bQ\big)\Big|^2\, 
\int_0\,d\ep\, 
\fract{\mathcal{N}_\text{IN}(\ep,\bQ)}{\big(\omega+\ep\big)^2}\,
\Bigg\},
\eea
where 
\bea
Z_\text{eq.} &=& 1 - 
\fract{(k_F a)^3}{8\pi^2}\,\int_0^2 \,Q\,dQ\,
\Bigg\{ \int_0^{Q(2-Q)}\,d\omega\,
\int_0 \fract{d\ep}{\pi}\, 
\fract{\Im\text{m} T\big(\ep-i0^+,\bQ\big)}{\big(\ep+\omega\big)^2}
\nonumber\\
&& + \int_0^{Q(2+Q)}\,d\omega\,
 \int_0 \fract{d\ep}{\pi}\, \fract{\Im\text{m} T\big( \! -\ep+i0^+,\bQ\big)}{\big(\ep+\omega\big)^2}\,
\Bigg]\Bigg\},\label{Z-eq}
\eea
\ew
is the equilibrium value at zero temperature\cite{Galitskii} and $a$ the lattice spacing. 
 
\subsection{Useful formulas}

Let us consider a function $F(Q,\omega_{\bk,\bQ})$ and define as 
$n_\bk$ the non-interacting momentum distribution. Because of momentum isotropy 
\[
f_\text{IN}(\bk) = \frac{1}{V}\sum_\bQ\,F(Q,\omega_{\bk,\bQ})\,n_\bk\,
n_{-\bk+\bQ},
\]
only depends on $|\bk|=k$. Therefore
\bea
f_\text{IN}(k_F) &=& \rho_0^{-1}\,\frac{1}{V}\sum_\bk\, f(\bk)\,
\delta\big(\epsilon_\bk\big) \nonumber \\
&=& \rho_0^{-1}\,
\frac{1}{V}\sum_\bQ\,\frac{1}{V}\sum_\bk\,\delta\big(\epsilon_\bk\big)\,F(Q,\omega_{\bk,\bQ})\,n_\bk\,
n_{-\bk+\bQ}\nonumber\\
&=&\rho_0^{-1}\,\frac{1}{V}\sum_\bQ\, 
\int_0 d\omega\, \rho_\text{IN}(\omega,\bQ)\, F(Q,\omega),\label{A1}
\eea
where $\rho_0$ is the non-interacting density of states at the 
Fermi energy and 
\bea
\rho_\text{IN}(\omega) &=& \frac{1}{V}\sum_\bk\,\delta\big(\epsilon_\bk\big)\,
\delta\Big(\omega-\fract{\omega_{\bk,\bQ}}{\ep_F}\Big)\,n_\bk\,
n_{-\bk+\bQ} \nonumber\\
&=& \fract{\rho_0}{4 Q}\;\theta\big(Q(2-Q)-\omega\big)\,
\theta(2-Q).\label{A2}
\eea
Seemingly, if 
\[
f_\text{OUT}(\bk) = \frac{1}{V}\sum_\bQ\,F(Q,\omega_{\bk,\bQ})\,
\big(1-n_\bk\big)\,
\big(1-n_{-\bk+\bQ}\big),
\]
then 
\be
f_\text{OUT}(k_F) = \rho_0^{-1}\,\frac{1}{V}\sum_\bQ\, 
\int_0 d\omega\, \rho_\text{OUT}(\omega,\bQ)\, F(Q,\omega),\label{A3}
\ee
where 
\bea
\rho_\text{OUT}(\omega) &=& \frac{1}{V}\sum_\bk\,\delta\big(\epsilon_\bk\big)\,
\delta\Big(\omega-\fract{\omega_{\bk,\bQ}}{\ep_F}\Big)\,n_\bk\,
n_{-\bk+\bQ} \nonumber \\
&=& \fract{\rho_0}{4 Q}\;\theta\big(Q(2+Q)-\omega\big).
\label{A4}
\eea
The above expressions are useful to evaluate the momentum distribution jump at $k_F$.


\end{document}